\documentclass{aa}  

\usepackage{graphicx}
\usepackage{subcaption}
\usepackage{caption}
\DeclareCaptionLabelFormat{cont}{#1~#2\alph{ContinuedFloat}}
\usepackage{txfonts}
\usepackage{xcolor}
\usepackage[flushleft]{threeparttable}
\usepackage{natbib}
\bibpunct{(}{)}{;}{a}{}{,} 

\begin{document}

   \title{Bright Lyman~$\rm \alpha$ emitters among {\it Spitzer} SMUVS galaxies in the MUSE/COSMOS field}

   \author{G. Rosani\inst{1}\fnmsep\thanks{Email: rosani@astro.rug.nl}\and
          G. B. Caminha\inst{1}\and
          K. I. Caputi\inst{1,2}\and
	  S. Deshmukh\inst{1}
          }

   \institute{Kapteyn Astronomical Institute, University of Groningen,
                   P.O. Box 800, 9700AV Groningen, The Netherlands
              \and
              Cosmic Dawn Center (DAWN), Niels Bohr Institute, University of Copenhagen,
                  Juliane Maries vej 30, DK-2100 Copenhagen, Denmark
             }

   \date{Received XXXX; accepted YYYY}

 
  \abstract
  {We search for the presence of bright Ly$\rm \alpha$ emitters among Spitzer SMUVS galaxies at $z>2.9$ making use of homogeneous MUSE spectroscopic data. Although these data only cover a small region of COSMOS, MUSE has the unique advantage of providing spectral information over the entire field, without the need of target pre-selection. This results in an unbiased detection of all the brightest Ly$\rm \alpha$ emitters among the SMUVS sources, which by design are stellar-mass selected galaxies. Within the studied area, $\rm \sim 14$\% of the SMUVS galaxies at $z>2.9$ have Ly$\rm \alpha$ fluxes $\rm F_\lambda\gtrsim 7\times 10^{-18}\, erg\, s^{-1}\, cm^{-2}$. These Ly$\rm \alpha$ emitters are characterized by three types of emission, 47\% show a single line profile, 19\% present a double peak or a blue bump and 31\% show a red tail. One object (3\%) shows both a blue bump and a red tail. We also investigate the spectral energy distribution (SED) properties of the SMUVS galaxies which are MUSE detected and which are not. After stellar-mass matching both populations, we find that the MUSE detected galaxies have generally lower extinction than SMUVS-only objects, while there is no clear intrinsic difference in the mass and age distributions of the two samples. For the MUSE-detected SMUVS galaxies, we compare the instantaneous SFR lower limit obtained from the Ly$\rm \alpha$ line with its past average derived from SED fitting, and find evidence for rejuvenation in some of our oldest objects. In addition, we study the spectra of those Ly$\rm \alpha$ emitters which are not detected in SMUVS in the same field. We find that the emission line profile shown are 67\% a single line, 3\% a blue bump and 20\% a red tail profile. The difference in profile distribution could be ascribed to the fainter Ly$\rm \alpha$ luminosities of the MUSE sources not detected in SMUVS and an intrinsically different mass distribution. Finally, we search for the presence of galaxy associations using the spectral redshifts. MUSE's integral coverage reveals that these associations are 20 times more likely than what is derived from all the other existing spectral data in COSMOS, which is biased by target pre-selection.}


   \maketitle
%

\section{Introduction}
\label{sec:intro}

The Lyman $\rm \alpha$ (Ly$\rm \alpha$) line contains important information about some of the main physical processes occurring in galaxies. Particularly, bright Ly$\rm \alpha$ emitters are tracers of the most prominent unobscured star formation activity at different cosmic times. The interpretation of the Ly$\rm \alpha$ line is however not trivial, because of its resonant nature, since it is easily scattered by the interstellar medium (ISM) on its way out of the galaxy. Furthermore, Ly$\rm \alpha$ photons are absorbed by dust and re-emitted at longer wavelengths, thus subtracting them from the line intensity. The kinematics of the gas also needs to be taken into account. All these processes give rise to different line profile shapes depending on the conditions surrounding the emitter galaxy \citep{2003ApJ...588...65S,2014A&A...565A...5K,2015ApJ...803....6M,2017arXiv170403416D,2017ApJ...849...82H,2017A&A...599A..28K,2018ApJ...852....9B,2018ApJ...862L..10E,2018arXiv181109630G,2018MNRAS.477.2098N,2018A&A...616A..60O,2018MNRAS.477.2817S,2018MNRAS.476L..15V,2019arXiv190105990K,2019arXiv190308593M,2019MNRAS.482.4553R,2019MNRAS.484...39S}.\par
It is possible to recover the original Ly$\rm \alpha$ flux, if information on the H$\rm \alpha$ line is available. There is no canonical conversion factor between Ly$\rm \alpha$ and H$\rm \alpha$ because secondary effects influence the conversion, but assuming case B recombination the values can reasonably range from $\approx 8$ \citep{2017arXiv170403416D} to $\sim 8.7$ \citep{1998ApJ...502L..99H}. When information on the H$\rm \alpha$ line is not present, we can rely on radiative transfer models exclusively treating the Ly$\rm \alpha$ line to try and recover the original emission from line shape fitting. Such models need to take the composition of the circumgalactic medium into account, the presence of dust, how dense the neutral hydrogen is, as well as the gas dynamics and the time evolution of the medium along with the star formation event. Models reproducing the shape of the Ly$\rm \alpha$ emission go from the early approach of \citet{1999MNRAS.309..332T} and \citet{2003ApJ...598..858M} to the more recent models by e.g. \citet{2008A&A...491...89V,2018MNRAS.478L..60V}, \citet{2017A&A...608A.139G} and \citet{2019arXiv190502480K}. Finally, selecting objects using the Ly$\rm \alpha$ line is a way to ensure that the more active star forming non-dusty galaxies are selected \citep{2019arXiv190509841Z}.\par
The use of {\it Spitzer} \citep{2004ApJS..154....1W} data in the 3.6/4.5 $\rm \mu$m bands allows us to access the red, flat part of the spectrum of high redshift galaxies. This results in a stellar mass selection, as the more luminous high-z galaxies at those wavelengths are typically the most massive. Such objects are interesting because they represent possible progenitors of today's most massive galaxies and possibly played an integral role in the peak of star formation history around $\rm z \sim\, 2$ \citep{2011MNRAS.413..162C,2018ApJ...864..166D,2018A&A...620A.198M}.\par
By combining both the more massive galaxies selected by {\it Spitzer} and the most prominent Ly$\rm \alpha$ emitters, we probe a very specific part of the galaxy evolution picture. We not only select the most massive objects at early times, but we also ensure that they are intensely star-forming and contain relatively little dust by detecting their Ly$\rm \alpha$ emission.\par
Previous studies combining Ly$\rm \alpha$ information with photometric SED fitting have been able to constrain stellar mass, age, star formation rate (SFR) and E(B-V) of Ly$\rm \alpha$ emitters among other properties. The general result of these studies is that Ly$\rm \alpha$ emitters are young, prominently star-forming, mostly unobscured intermediate-mass galaxies. Some of these studies also find evidence for more massive objects being present among the Ly$\rm \alpha$ population. Moreover, the age distribution of the Ly$\rm \alpha$ emitters presents in some cases an age bimodality, for which a significant fraction of the galaxies studied is not young, but has ages around 1 Gyr \citep{2008ApJ...674...70L,2009ApJ...691..465F,2010MNRAS.402.1580O,2009A&A...494..553P,2010A&A...514A..64P,2010ApJ...720.1016Y,2011ApJ...733..114G,2012ApJ...751L..26A,2012ApJ...760..128M,2013MNRAS.429..302C,2014MNRAS.439..446M,2017A&A...608A.123D,2018ApJ...864..145H,2019arXiv190308593M}.\par
Among the best studied blank areas of the sky is the COSMOS field \citep{2007ApJS..172....1S}, for which a wide range of homogeneous and deep datasets are available. One of the surveys spanning part of the field is the {\it Spitzer Matching Survey of the UltraVISTA ultra-deep Stripes} (SMUVS, \citealt{2018ApJS..237...39A}). The SMUVS survey combines {\it Spitzer} 3.6 and 4.5 $\rm \mu$m data with 26 complementary photometric bands to infer the redshift and physical properties of its galaxies. Most importantly, it uses the {\it Spitzer} InfraRed Array Camera (IRAC) data \citep{2004ApJS..154...10F} to gain access to the flat part of the continuum of high redshift objects. The SMUVS survey has already produced a number of studies of galaxy properties at redshift $\rm z>2$ \citep{2017ApJ...849...45C,2018ApJ...853...69C,2019ApJ...874..114C,2018ApJ...864..166D}.\par
The Multi Unit Spectroscopic Explorer spectrograph (MUSE, \citealt{2010SPIE.7735E..08B}) provides a powerful tool to expand the existing analysis of any part of the sky. Being an integral field unit (IFU) spectrograph, it allows to take the spectra of a $\sim 1'\times 1'$ portion of the sky with no source pre-selection. Furthermore, new sources can be discovered serendipitously and biases in the galaxy sample, inherent to slit spectroscopy, can be avoided.\par
As we aim to combine MUSE and SMUVS, and given the wavelength range covered by MUSE, the spectral feature available for us to study in objects above $\rm z\sim 3$ is the Ly$\rm \alpha$ emission. This line emission is visible in the MUSE spectral range for objects at redshifts $\rm 2.9 \lesssim z \lesssim 6.6$.\par
The scope of this paper is to analyze the physical properties of the more prominent Ly$\rm \alpha$ emitters detected in MUSE. We will thus use homogeneous MUSE observations in sub-regions of the COSMOS field of view to spectroscopically confirm SMUVS sources in an area of $\rm \approx 20.79\, arcmin^2$. We will give special attention to sources in the redshift range $\rm 2.9 \le\, z\le\, 6.6$, where we can obtain secure confirmation from the Ly$\rm \alpha$ line with MUSE, and study its physical properties using the rich broad band photometry from SMUVS, also comparing with the sample of non-Ly$\rm \alpha$ emitters at the same redshift range. Furthermore, we also list new sources detected by a blind search performed in the MUSE pointings and test our spectroscopic sample for possible physical associations.\par
In Section~\ref{sec:data} we describe the data we used, in Section~\ref{sec:results} we outline our results and in Section~\ref{sec:conclu} we discuss our results and draw our conclusion. Throughout this paper, we adopt a flat $\rm \Lambda$CDM cosmology with $\rm H_o=69.6\, km\, s^{-1}\, Mpc^{-1}$, $\rm \Omega_m=0.286$ and $\rm \Omega_\Lambda=0.714$.


\section{Data} 
\label{sec:data} 

The COSMOS field is one of the most observed regions of the sky with numerous ancillary data sets obtained from ground and space based observatories. In this section we describe only the datasets that we have used for our sample selection. 
 
\subsection{SMUVS sources} 
\label{sec:smuvs_photometry}

We use the version of the SMUVS catalog \citep{2018ApJS..237...39A} presented in \citet{2018ApJ...864..166D} to select part of our sources. We focus on the area of the MUSE/COSMOS GTO field. SMUVS is a {\em Spitzer Space Telescope} Exploration Science Program which combines observations in the IRAC \citep{2004ApJS..154...10F} $\rm 3.6\mu$m and the $\rm 4.5\mu$m bands, taken over $\sim 0.66 \, \rm deg^2$ of the COSMOS field. The area observed by SMUVS overlaps with the three UltraVISTA ultra-deep stripes \citep{2012A&A...544A.156M} and with the COSMOS deepest optical coverage of the Subaru telescope \citep{2007ApJS..172....9T}.\par 
The SMUVS source detection is a double-selection in the H$\rm K_s$ average stack maps constructed using data from the UltraVISTA third data release and in the 3.6/4.5 $\rm \mu$m IRAC bands. As described in \citet{2018ApJ...864..166D}, SExtractor \citep{1996A&AS..117..393B} is applied on the H$\rm K_s$ UltraVISTA maps to select sources that will then be used as priors to perform a point spread function (PSF) fitting on the IRAC images to finalize the selection. The SMUVS catalog includes multi-wavelength photometric data available for COSMOS in 26 bands, from the U through the UltraVISTA $\rm K_s$ band. All this photometric data, along with the IRAC photometry, has been used to perform the spectral energy distribution (SED) fitting and derive physical properties for about 300,000 galaxies \citep{2018ApJ...864..166D}. In this paper we consider the $\rm \sim 3,000$ SMUVS sources that lie on the $20.79 \, \rm arcmin^2$ area of the MUSE COSMOS/GTO program.\par 

\subsection{MUSE spectroscopy} 
\label{sec:muse_spectroscopy} 

In this work we analyze archival data from MUSE \citep{2010SPIE.7735E..08B} in the COSMOS field \citep{2007ApJS..172....1S}, over an area of $20.79\, \rm arcmin^2$  embedded in the SMUVS footprint \citep{2018ApJS..237...39A}. MUSE is one of the latest spectrographs mounted on the Very Large Telescope and offers integral field spectroscopy over an entire $ 1 \,\rm arcmin^2$ field of view, providing a spectrum for each  $\rm 0.2 \times 0.2 \, \rm arcsec^2$ pixel element. Therefore, the observations do not require pre-selection of targets and thanks to its small pixel size, the source separation is limited only by observational conditions. MUSE covers the wavelength range $\rm 4750-9350\, \AA$ with a spectral bin of $\rm 1.25\, \AA/pixel$, resulting in an average resolving power $\rm R \approx 3000$ at $\lambda \sim 7500\, \AA$ and an almost constant resolution of $\rm \Delta\lambda \approx 2.4 \, \AA$.\par
We made use of a homogeneous data set obtained by the  MUSE consortium under the Guaranteed Time Programme IDs 095.A-0240, 096.A-0090, 097.A-0160 and 098.A-0017 (P.I.: L. Wisotzki), as part of the so called MUSE-Wide survey \citep{2017A&A...606A..12H, 2017MNRAS.471.3186D}. The observations consist in 23 different MUSE pointings with one hour of exposure time, of which 21 form a contiguous area in a $3 \times 7$ mosaic, and the remaining two are located in a region $\rm \approx 5 \, arcmin$ apart. The data acquisition was carried out under fair observational conditions with a median seeing of $1\arcsec.10$, from the DIMM station measurements, and $\approx 17\%$ of the exposures have seeing higher than $1\arcsec.5$. The field of view was chosen in order to overlap with deep HST imaging from the CANDELS/COSMOS survey \citep{2011ApJS..197...35G,2011ApJS..197...36K}, maximizing the amount of photometric information we have on our objects. Additional MUSE pointings from different GO and GTO programs overlapping with the SMUVS field are publicly available. However, we do not consider them here as we aim to work with a data set of homogeneous depth for the seek of clarity in our results/conclusions.\par
We retrieved the MUSE raw exposures and calibration files from the ESO archive and used the standard reduction pipeline version 2.0.3 \citep{2006NewAR..50..405W,2012SPIE.8451E..0BW,2014ASPC..485..451W} in combination with the MUSE Python Data Analysis Framework \citep[MPDAF version 2.3,][]{2016ascl.soft11003B, 2017arXiv171003554P} and the Zurich Atmosphere Purge \citep[ZAP version 2.1,][]{2016MNRAS.458.3210S} to create the final data cubes.\par
Finally, we corrected the WCS coordinates by using SExtractor to identify the centroids of the brighter objects in the MUSE white images and the CANDELS HST F160W image. We verified that MUSE shows an average offset with respect to HST of $\rm 0.141''$ with standard deviation $\rm 0.110''$, while the offset with the reported SMUVS coordinates is $\rm 0.196\pm 0.123''$.  We corrected the MUSE coordinates taking the HST centroids as reference and note that the offsets between the different catalogs are always smaller than the MUSE pixel scale (0.2 arcsec).

\section{Results}
\label{sec:results}
\subsection{SMUVS sources in the MUSE/COSMOS GTO fields}
\label{sec:SMUVSMUSE}
\begin{figure}
\centering
\includegraphics[scale=0.4]{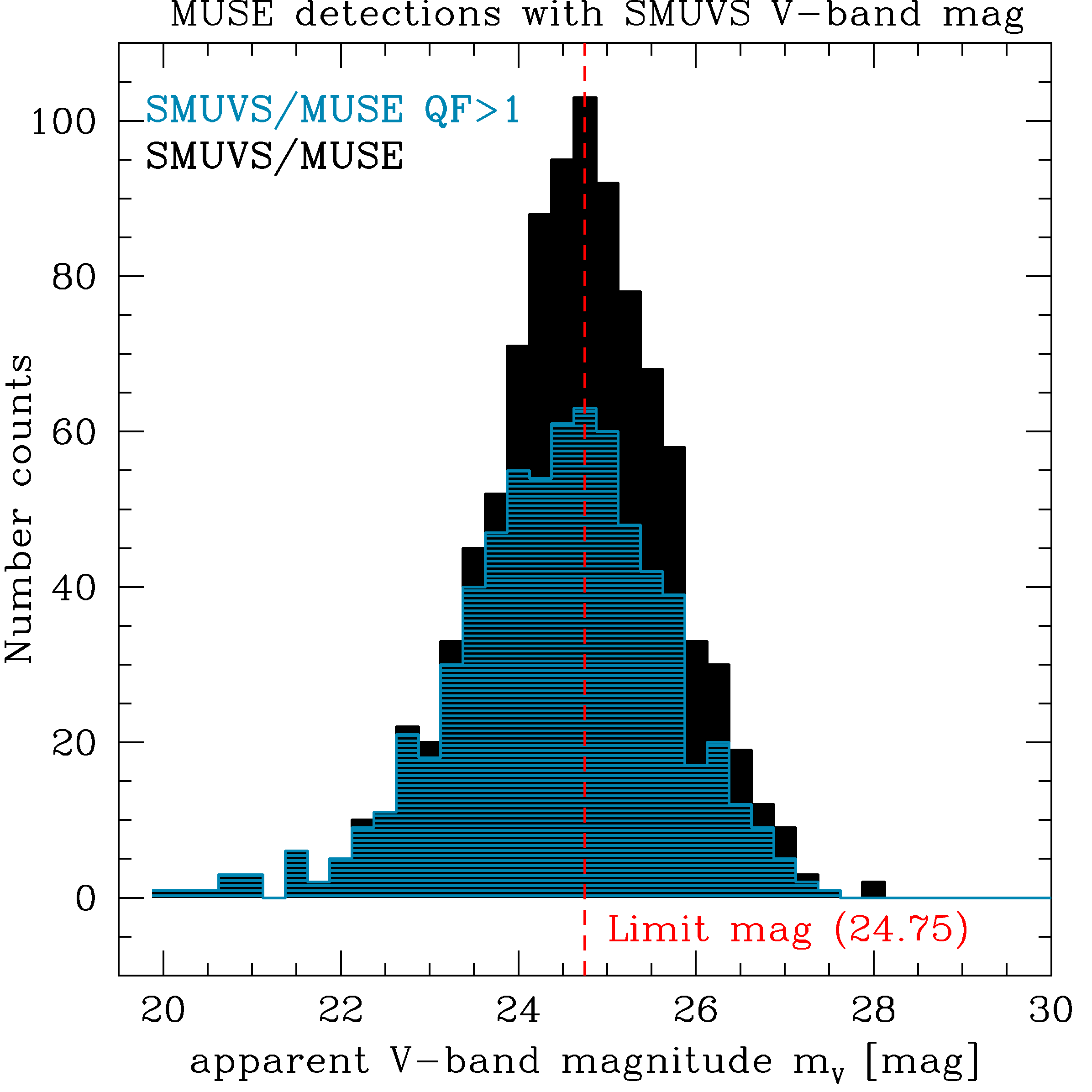}
\caption{Histogram of the V-band magnitudes assigned by SMUVS to our MUSE detected sources. The peak of the distribution shows the magnitude after which the number counts of MUSE drop significantly.}
\label{fig:histoLimMag}
\end{figure}
\begin{figure}
\centering
\includegraphics[scale=0.45]{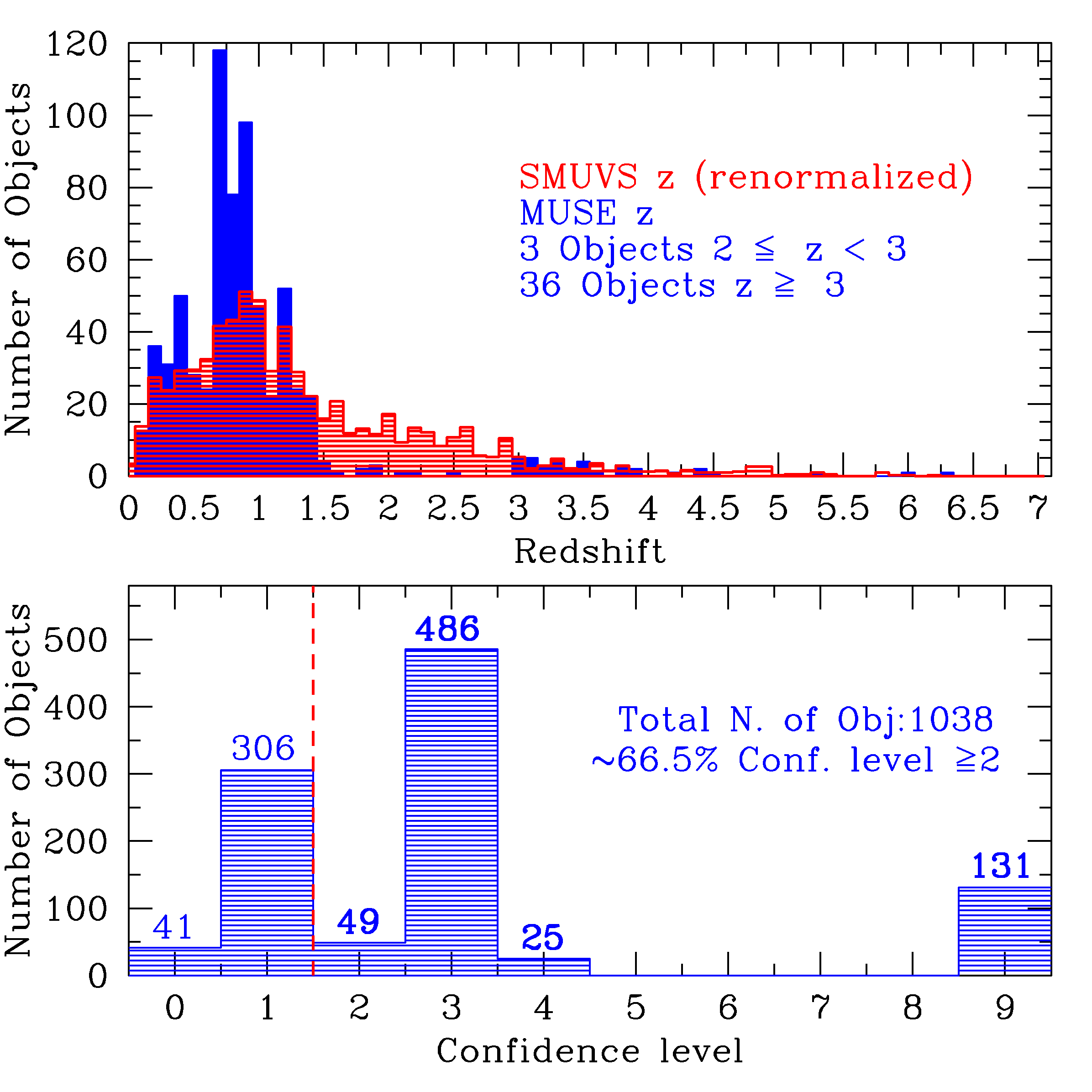}
\caption{{\em Upper panel:} redshift distribution of our sample of SMUVS/MUSE sources with spectral $\rm QF \ge 2$ in blue. These objects constitute $\rm \sim 66.5 \%$  of the 1038 SMUVS objects identified with MUSE, and $\sim 23\%$ of all the SMUVS objects in the COSMOS/MUSE GTO field. The distribution of all the SMUVS redshifts in our MUSE fields is drawn in red and renormalized to the number of MUSE sources detected with $\rm QF \ge 2$. {\em Lower panel:} distribution of spectral QF for the 1038 MUSE detected sources. The dashed line indicates the boundary between a secure and an uncertain redshift measurement. Quality flag 0 was assigned to all galactic stars, independently of their actual spectral quality. }
\label{fig:histoQF}
\end{figure}
\begin{table}
\caption{SMUVS high-redshift ($z\geq2$) sources identified in the COSMOS/MUSE GTO field. The SMUVS ID, position on the sky, measured spectroscopic redshift and quality flags are listed.} 
\label{table:HZsource}      
\begin{threeparttable}
\centering                          
\small
\begin{tabular}{l|c c|c|c}        
\hline\hline                 
Obj. ID & RA & Dec & $z_{spec}$ & QF\\    
\hline                        
  73023 & 150.1778406 & 2.1941108 & 3.8686 & 2\\
  73055 & 150.1044581 & 2.1943565 & 3.7710 & 9\\
  73162 & 150.1868076 & 2.1950781 & 4.4239 & 9\\
  73174 & 150.1807239 & 2.1949942 & 3.3375 & 3\\
  73452 & 150.0935903 & 2.1970951 & 3.0377 & 3\\
  73503 & 150.1657897 & 2.1971041 & 3.2774 & 9\\
  73761\tnote{1} & 150.0973771 & 2.1988546 & 3.0747 & 3\\
  73993 & 150.1835347 & 2.2011801 & 3.2854 & 9\\
  74055 & 150.1698170 & 2.2012076 & 3.1317 & 9\\
  74237 & 150.1139483 & 2.2015449 & 3.4416 & 9\\
  74717 & 150.1050908 & 2.2068064 & 3.6090 & 9\\
  74990 & 150.0925832 & 2.2078347 & 3.5265 & 3\\
  75041 & 150.1815360 & 2.2085007 & 4.4443 & 3\\
  75190 & 150.1038713 & 2.2103898 & 3.2407 & 9\\
  75249 & 150.1209257 & 2.2102457 & 3.2212 & 9\\
  75267 & 150.1876116 & 2.2097678 & 2.4796 & 2\\
  75288 & 150.1393225 & 2.2109674 & 5.2944 & 3\\
  75461 & 150.1469868 & 2.2122848 & 3.4608 & 3\\
  75570 & 150.1181685 & 2.2128694 & 3.0031 & 9\\
  75825 & 150.1249549 & 2.2145933 & 5.9886 & 3\\
  76037 & 150.1601037 & 2.2163066 & 3.0060 & 3\\
  76038 & 150.0944809 & 2.2157858 & 3.3990 & 9\\
  76101 & 150.1703151 & 2.2160544 & 3.5259 & 3\\
  76321 & 150.1410657 & 2.2179212 & 3.9089 & 3\\
  76802 & 150.1064151 & 2.2215919 & 6.3044 & 9\\
  76829 & 150.1926599 & 2.2198279 & 3.0900 & 3\\
  76877 & 150.1659591 & 2.2216971 & 3.0913 & 3\\
  77599 & 150.1563335 & 2.2255982 & 3.1819 & 3\\
  77637 & 150.1334882 & 2.2276703 & 4.1650 & 3\\
  78106 & 150.1175785 & 2.2315227 & 2.9722 & 3\\
  78164 & 150.1058933 & 2.2303365 & 3.0097 & 3\\
  78359 & 150.1800388 & 2.2312780 & 2.1458 & 3\\
  78448 & 150.1866714 & 2.2319585 & 2.1723 & 2\\
  78588 & 150.1167228 & 2.2351184 & 3.8206 & 9\\
  78635 & 150.0881948 & 2.2344639 & 3.4881 & 3\\
  78718 & 150.1097010 & 2.2362429 & 3.7711 & 3\\
  90896 & 150.1160814 & 2.3273753 & 3.2644 & 9\\
  91354 & 150.1226590 & 2.3311551 & 3.0957 & 9\\
  91380 & 150.1274174 & 2.3309118 & 4.4672 & 9\\
\hline                                   
\end{tabular}
\begin{tablenotes}
\item[1] $z_{MOSFIRE}=3.0768$ \citep{2015ApJS..218...15K}.
\end{tablenotes}
\end{threeparttable}
\end{table}
We searched for detections of the SMUVS sources in the $20.79 \, \rm arcmin^2$ covered by the MUSE datacubes, and measure their redshifts from emission/absorption lines in their spectra. We consider a SMUVS counterpart detected in MUSE when the MUSE emission arises within $\rm 1''$ from the SMUVS source centroid. Furthermore, we use HST images from CANDELS to verify possible contamination from nearby sources. We find that the MUSE emission can always be univocally assigned to one source and discuss the implications of source contamination on the photometry of our sample in more detail in Sect.~\ref{sec:inSMUVS}.\par
As explained above, the advantage of analyzing the MUSE data with respect to any other spectroscopic dataset in COSMOS is that MUSE does not require a source pre-selection and, thus, the identification completeness is solely governed by the spectroscopic depth. Fig.~\ref{fig:histoLimMag} shows that if we translate the depth to a V-band magnitude, the number counts in the MUSE pointings start to drop significantly after a magnitude of $\rm \sim 24.75$. We consider this limiting magnitude to compare the depth of the MUSE pointings to the depth of the SMUVS survey.\par
Out of 2997 SMUVS objects present in the area covered by MUSE, we managed to successfully identify 1038 objects spectroscopically. All redshifts have been measured and agreed upon by two independent observers, following the work philosophy adopted in, e.g., the zCOSMOS spectroscopic survey \citep{2007ApJS..172...70L}.\par
Also similarly to zCOSMOS, we classify the quality of our spectra by applying the following quality flags (QF):
\begin{itemize}
\item -99: non-detection;
\item 0: galactic stars, independently of the spectral quality;
\item 1: redshift measurement is only tentative;
\item 2: relatively secure redshift measurement, with the spectrum showing faint line(s) and/or a continuum, for which the redshift is likely to be correct;
\item 3: very secure redshift measurement, typically based on more than one emission line and/or a clear continuum with absorption lines;
\item 4: text-book spectrum with emission and absorption lines, and a very clear continuum;
\item 9: redshift based on a single but clearly detected emission line, for which we are unsure about its identification. In these cases, a few alternative spectroscopic redshift values are possible for the source. 
\end{itemize}
The MUSE detection rate for the whole SMUVS sample is $\rm \sim 35\%$ (=1038/2997), considering all detections regardless of their QF.  Among these objects, a total of 691 have a spectroscopic redshift measurement with QF$\geq2$, i.e., $\sim 23\%$ of all the SMUVS sample in the COSMOS/MUSE GTO field, with the following distribution: 49 galaxies have QF=2;  486 have QF=3; 25 have  QF=4; and 131 are classified with QF=9, as shown in the lower panel of Fig.~\ref{fig:histoQF}. Furthermore, we also detected 41 stars (QF=0) and 306 galaxies for which the MUSE data quality is not good enough to constrain the redshift of the object with a high enough confidence level (QF=1). The remaining 1959 SMUVS objects are non-detections in MUSE.\par
The upper panel of Fig.~\ref{fig:histoQF} shows the redshift distribution of our SMUVS/MUSE sources in blue and the $\rm z_{phot}$ distribution of all the SMUVS sources in the COSMOS/GTO field in red. The SMUVS histogram has been renormalized to match the SMUVS/MUSE sample numbers. As is evident, most of the SMUVS/MUSE detections are located at low redshifts with two overdensities at $\rm z \sim 0.7$ and $\rm z \sim 0.9$, which belong to previously identified large-scale structures in the COSMOS field \citep{2005A&A...439..845L,2010ApJ...708..505K}. Since the MUSE data is shallow, it is natural that we only see the brighter sources at higher redshifts. We also can see that the MUSE detections clearly identify the overdensities, and favour redshifts where strong emission lines are present in the MUSE wavelength range (i.e. [OII] between $\rm \sim 0.2$ and $\rm \sim 1.4$ and Ly$\rm \alpha$ above $\rm z=2.9$). This is in contrast with the photometric redshift distribution, whose intrinsic dispersion smooths out the peaks, making it less suitable to identify preferential redshifts.\par
Among the SMUVS sources with MUSE QF $\ge 2$ there are 39 with spectroscopic redshift $z_{\rm spec} \geq 2$. These include three sources in the redshift range $\rm 2 \le\, z < 3$ and 36 with redshifts $\rm z \ge 3$. The $\rm 2 \le\, z < 3$ sources consist of two bright galaxies with absorption lines and one AGN with broad $\rm CIV$ and $\rm [CIII]$ emission, the $\rm z\ge 3$ sources are prominent Ly$\rm \alpha$ emitters. If we compare the blue and red histograms in Fig.~\ref{fig:histoQF}, we can see that the MUSE incidence is comparable to SMUVS until $\rm z>4$. For the higher redshift, it is the SMUVS relative incidence that is more pronounced.\par
At $\rm 1.5 \lesssim \, z \lesssim 3$ the number of identified SMUVS/MUSE objects is drastically lower than at higher redshifts. This is the so-called ``redshift desert'', where no strong nebular emission line falls into the wavelength range covered by MUSE and thus makes detection particularly difficult. This is clearly illustrated if we look at the number of sources identified in SMUVS in the same redshift range (see red histogram in Fig.~\ref{fig:histoQF}).\par
Table~\ref{table:HZsource} shows the results of our redshift measurements for our high-redshift ($\rm z_{spec} \geq2$) sample. We list the SMUVS ID and position of the objects, as well as the spectroscopic redshifts measured and the quality flag assigned to their spectra. There are three sources with $\rm QF=2$, additional 20 sources with $\rm QF=3$, and 16 sources with $\rm QF=9$. For some of the latter, the ambiguity in the single emission line identification could be solved via the available photometric redshifts of these sources \citep{2018ApJ...864..166D},  as will be discussed in Sect.~\ref{sec:inSMUVS}. We also notice that one of our objects (ID: \#73761) has a previous spectroscopic redshift identification obtained with MOSFIRE on the Keck Telescope \citep{2015ApJS..218...15K}, and that our own redshift is in good agreement with this previous value ($\rm z_{MOSFIRE}=3.0768$). All the remaining spectroscopic redshifts listed here are new, i.e. they are not present in the existing spectroscopic catalogs for the COSMOS field. \par

\subsection{The MUSE spectra}
\label{sec:MUSEspec}
\begin{table*}
\caption{Measured Ly$\rm \alpha$ fluxes and luminosities for our SMUVS/MUSE sources. The ID, center of both the main and secondary peak, the measured flux from the fit and the Ly$\rm \alpha$ luminosity are reported.}
\label{table:FluxLyA}
\centering
\small
\begin{tabular}{|c|c|c|c|c|}
\hline\hline                 
SMUVS ID & Line Center (Main) & Line Center (Sec.) & $\rm F_{fit}$ & $\rm L_{fit}$ \\
\#&$\rm [\AA]$&$\rm [\AA]$&$\rm [10^{-18}\, erg\, s^{-1}\, cm^{-2}]$&$\rm [10^{42}\, erg\, s^{-1}]$\\
\hline
73023&5917.9 $\pm$ 0.6&5920.4 $\pm$ 1.6&34.98 $\pm$ 4.98&5.18 $\pm$ 0.74\\
73055&5799.2 $\pm$ 0.5&-&13.42 $\pm$ 2.03&1.87 $\pm$ 0.28\\
73162&6594.1 $\pm$ 0.2&-&7.27 $\pm$ 0.85&1.48 $\pm$ 0.17\\
73174&5272.9 $\pm$ 0.2&-&37.46 $\pm$ 2.67&3.89 $\pm$ 0.28\\
73452&4909.5 $\pm$ 0.2&-&49.81 $\pm$ 3.57&4.13 $\pm$ 0.30\\
73503&5201.2 $\pm$ 0.6&5192.7 $\pm$ 0.8&46.32 $\pm$ 5.21&4.61 $\pm$ 0.52\\
73761&4958.2 $\pm$ 0.2&4950.0 $\pm$ 0.2&240.86 $\pm$ 8.01&20.56 $\pm$ 0.68\\
73993&5210.7 $\pm$ 0.6&-&13.76 $\pm$ 2.33&1.38 $\pm$ 0.23\\
74055&5024.1 $\pm$ 0.6&-&16.10 $\pm$ 3.02&1.44 $\pm$ 0.27\\
74237&5399.2 $\pm$ 0.4&5405.9 $\pm$ 0.6&32.37 $\pm$ 3.42&3.62 $\pm$ 0.38\\
74717&5604.2 $\pm$ 0.4&-&18.93 $\pm$ 2.78&2.37 $\pm$ 0.35\\
74990&5508.9 $\pm$ 0.2&-&46.55 $\pm$ 3.94&5.52 $\pm$ 0.47\\
75041&6618.6 $\pm$ 0.3&6610.5 $\pm$ 0.8, 6623.4 $\pm$ 3.9&17.11 $\pm$ 3.32&3.52 $\pm$ 0.68\\
75190&5156.4 $\pm$ 0.5&5145.2 $\pm$ 0.8&37.35 $\pm$ 5.16&3.61 $\pm$ 0.50\\
75249&5131.0 $\pm$ 1.0&-&17.94 $\pm$ 3.65&1.71 $\pm$ 0.35\\
75288&7652.2 $\pm$ 0.9&7656.8 $\pm$ 2.0&22.52 $\pm$ 4.55&7.00 $\pm$ 1.42\\
75461&5422.0 $\pm$ 0.3&5423.5 $\pm$ 1.1&21.55 $\pm$ 3.29&2.44 $\pm$ 0.37\\
75570&4866.6 $\pm$ 0.2&-&68.61 $\pm$ 4.62&5.53 $\pm$ 0.37\\
75825&8495.8 $\pm$ 0.2&8500.2 $\pm$ 0.3&43.01 $\pm$ 3.23&17.82 $\pm$ 1.34\\
76037&4868.8 $\pm$ 0.5&4873.1 $\pm$ 3.9&26.42 $\pm$ 10.52&2.13 $\pm$ 0.85\\
76038&5347.7 $\pm$ 0.8&-&16.98 $\pm$ 2.96&1.85 $\pm$ 0.32\\
76101&5501.9 $\pm$ 0.3&-&11.91 $\pm$ 1.58&1.41 $\pm$ 0.19\\
76321&5967.2 $\pm$ 0.2&5971.0 $\pm$ 2.4&50.53 $\pm$ 8.81&7.67 $\pm$ 1.34\\
76802&8879.8 $\pm$ 0.3&-&24.76 $\pm$ 3.13&11.56 $\pm$ 1.46\\
76829&4968.7 $\pm$ 0.5&4975.9 $\pm$ 0.9&32.19 $\pm$ 4.05&2.78 $\pm$ 0.35\\
76877&4973.4 $\pm$ 0.3&-&35.59 $\pm$ 3.44&3.08 $\pm$ 0.30\\
77599&5083.7 $\pm$ 0.2&-&64.82 $\pm$ 6.18&6.01 $\pm$ 0.57\\
77637&6278.8 $\pm$ 0.3&6284.5 $\pm$ 2.1&26.72 $\pm$ 3.46&4.71 $\pm$ 0.61\\
78106&4829.3 $\pm$ 0.2&-&48.94 $\pm$ 4.58&3.85 $\pm$ 0.36\\
78164&4875.8 $\pm$ 0.3&4883.2 $\pm$ 0.7&59.26 $\pm$ 6.23&4.80 $\pm$ 0.50\\
78588&5861.3 $\pm$ 0.7&5864.3 $\pm$ 1.2&39.44 $\pm$ 8.04&5.66 $\pm$ 1.15\\
78635&5456.2 $\pm$ 0.2&-&70.35 $\pm$ 4.68&8.13 $\pm$ 0.54\\
78718&5800.5 $\pm$ 0.2&-&121.65 $\pm$ 7.48&16.94 $\pm$ 1.04\\
90896&5185.3 $\pm$ 0.7&-&24.07 $\pm$ 3.85&2.37 $\pm$ 0.38\\
91354&4979.2 $\pm$ 0.2&4972.6 $\pm$ 0.4&40.92 $\pm$ 3.37&3.55 $\pm$ 0.29\\
91380&6646.6 $\pm$ 0.5&-&38.46 $\pm$ 4.09&8.02 $\pm$ 0.85\\
\hline
\end{tabular}
\end{table*}

\begin{figure*}[!ht]
\centering
\includegraphics[scale=0.18]{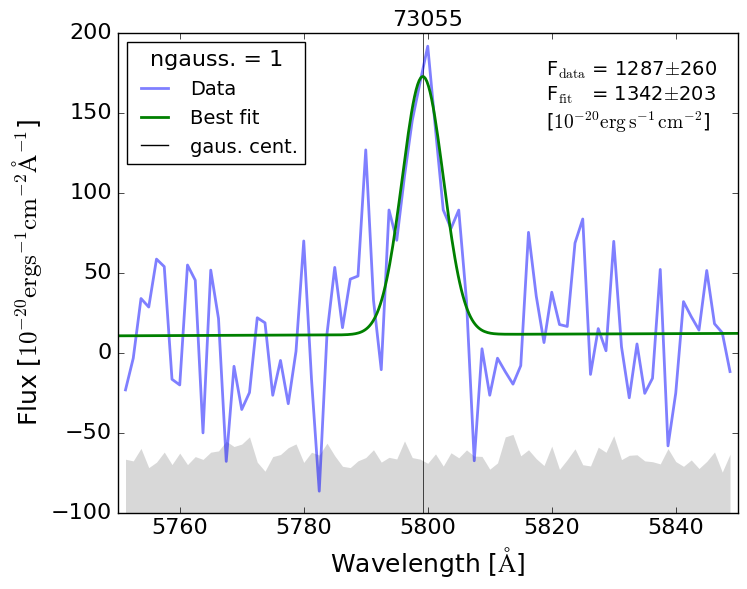}
\includegraphics[scale=0.18]{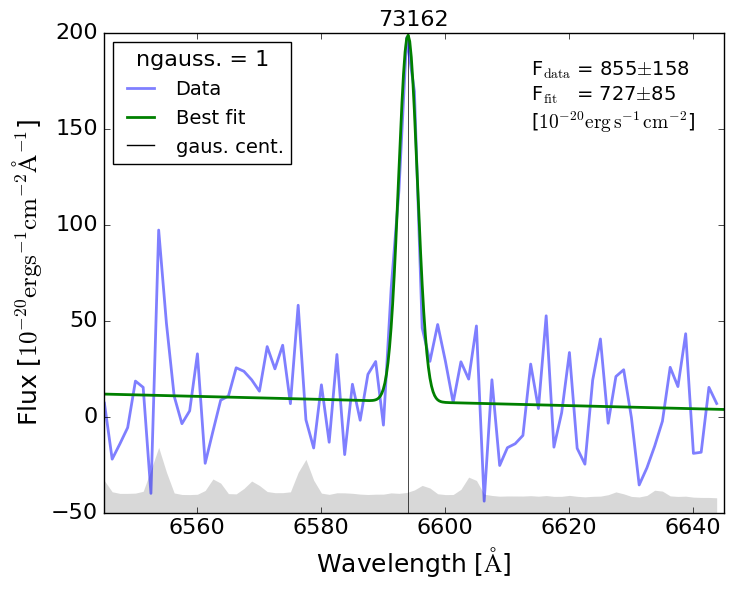}
\includegraphics[scale=0.18]{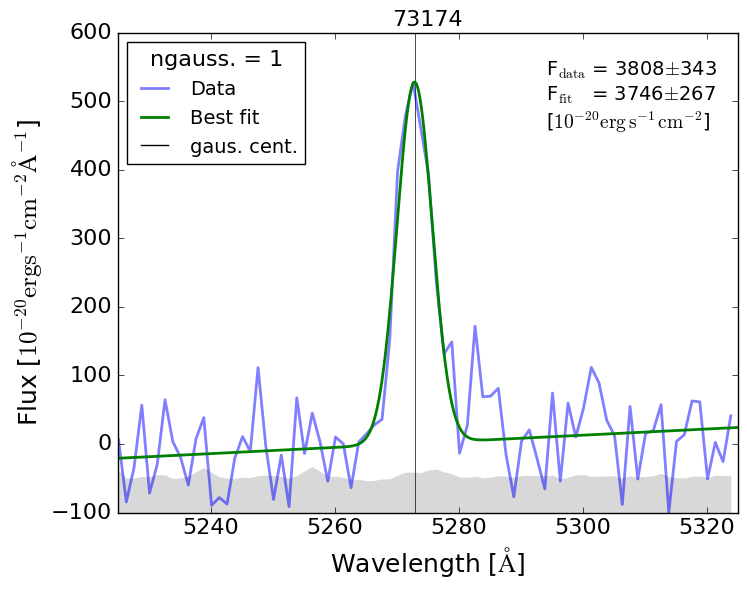}
\includegraphics[scale=0.18]{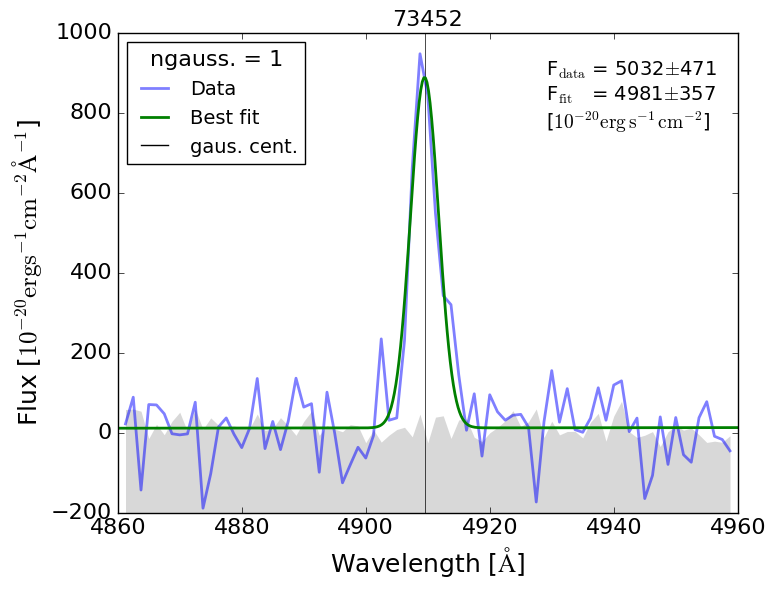}
\includegraphics[scale=0.18]{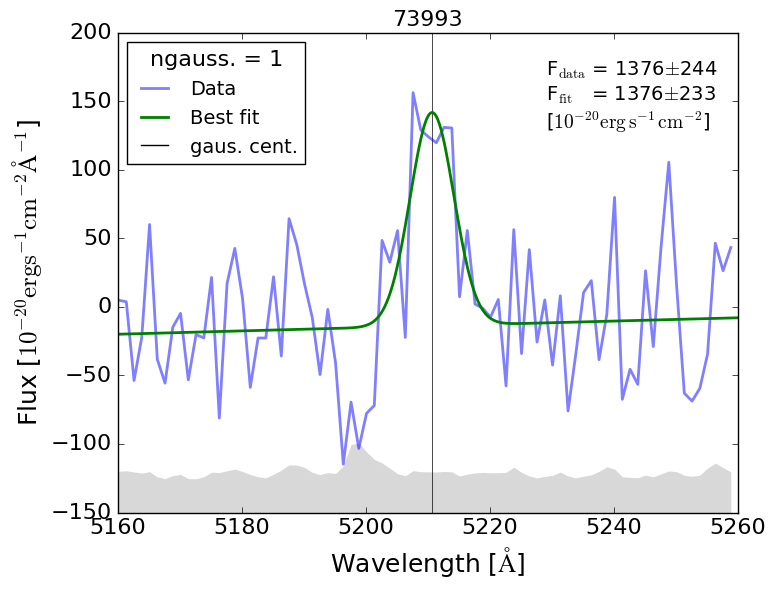}
\includegraphics[scale=0.18]{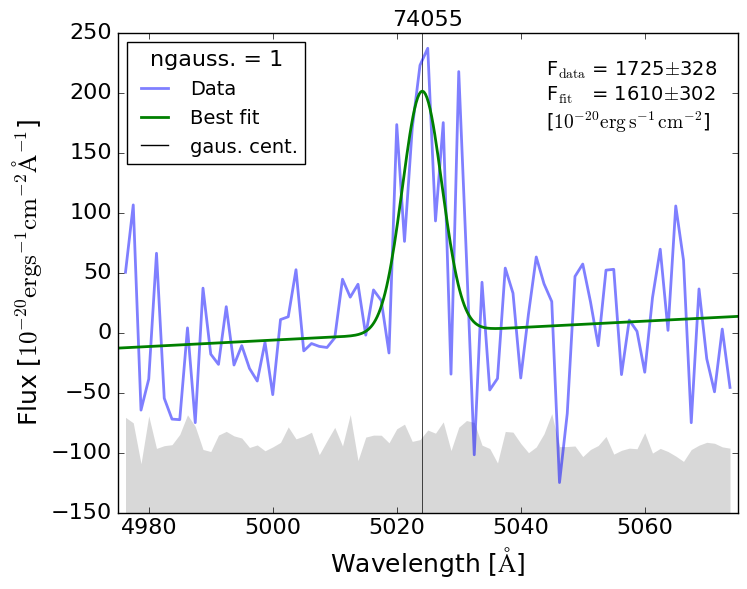}
\includegraphics[scale=0.18]{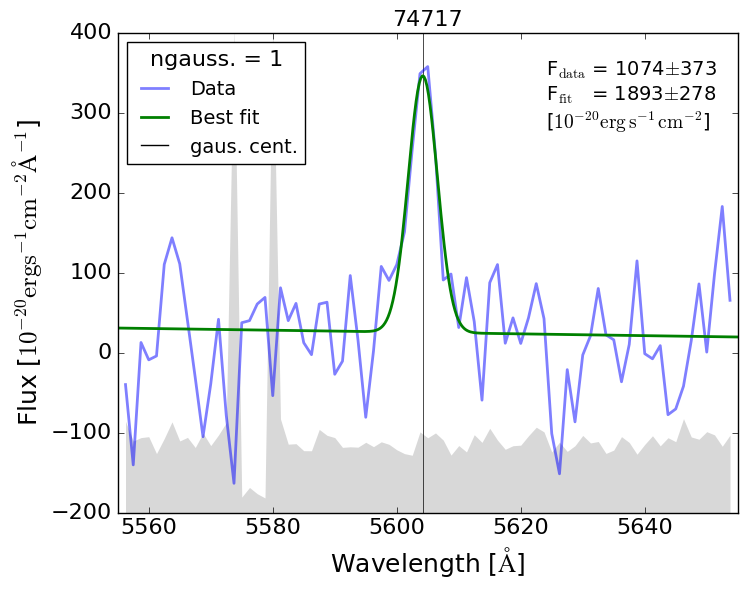}
\includegraphics[scale=0.18]{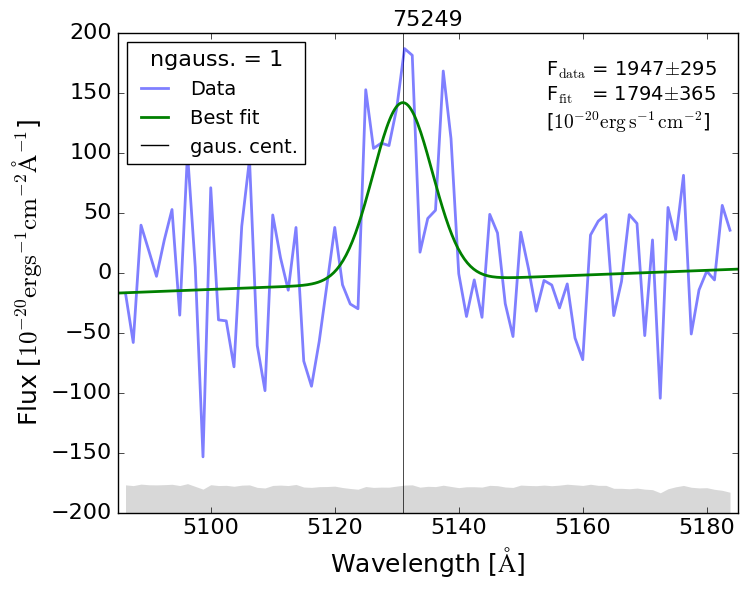}
\includegraphics[scale=0.18]{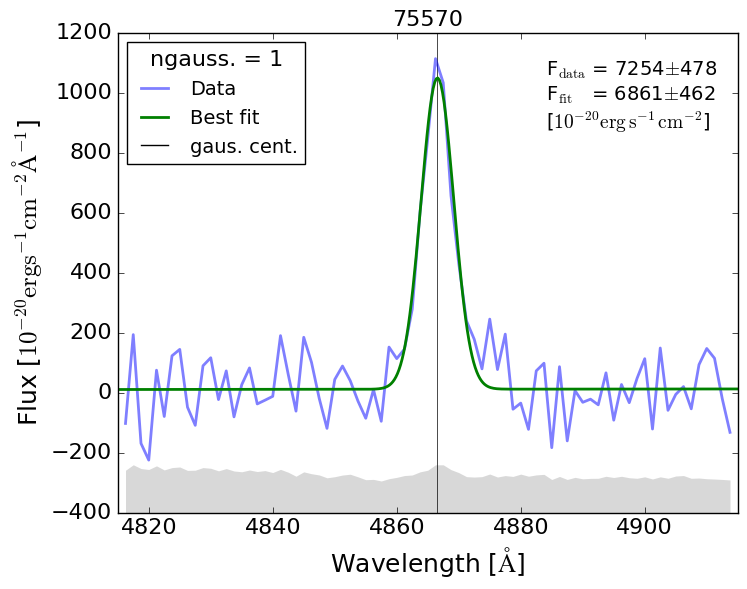}
\includegraphics[scale=0.18]{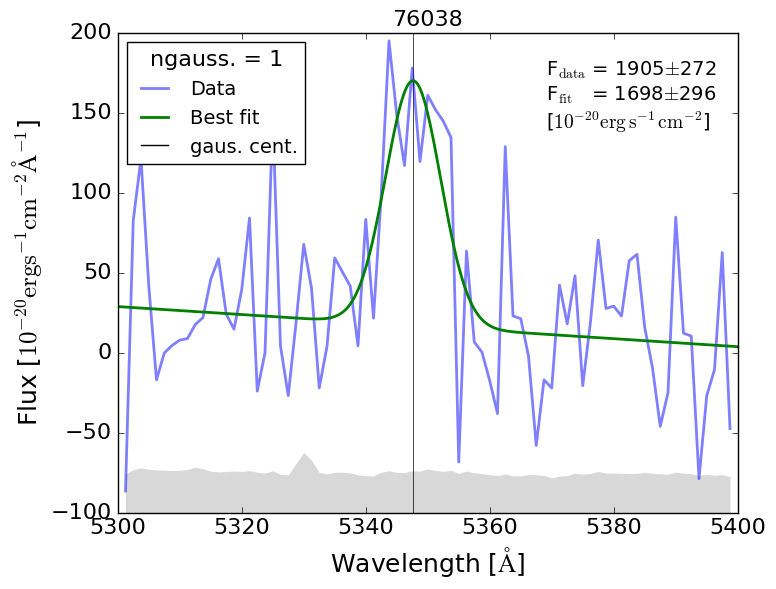}
\includegraphics[scale=0.18]{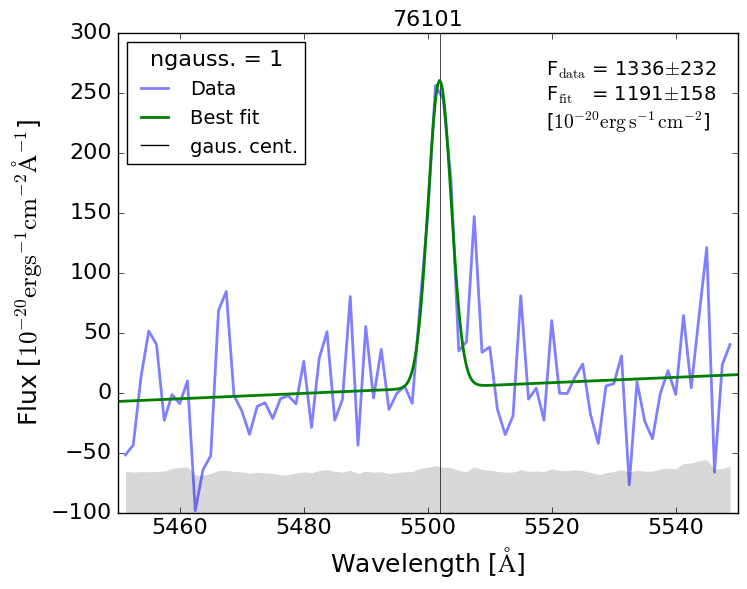}
\includegraphics[scale=0.18]{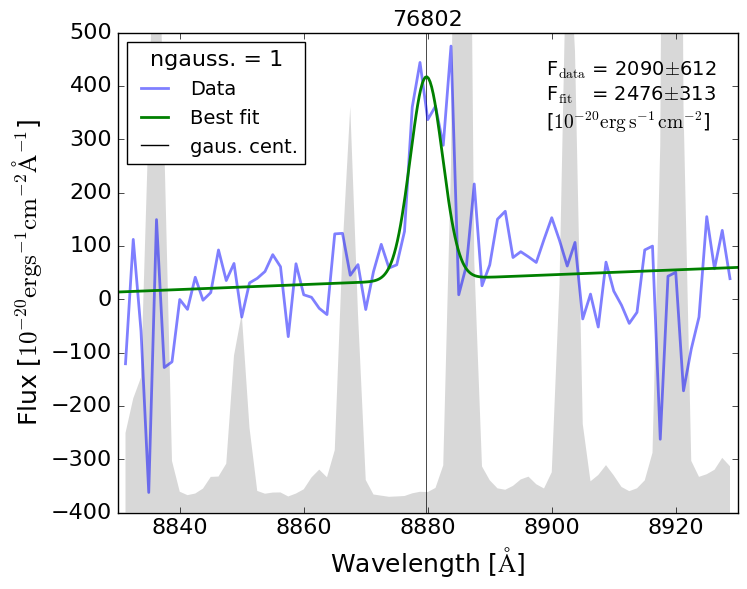}
\includegraphics[scale=0.18]{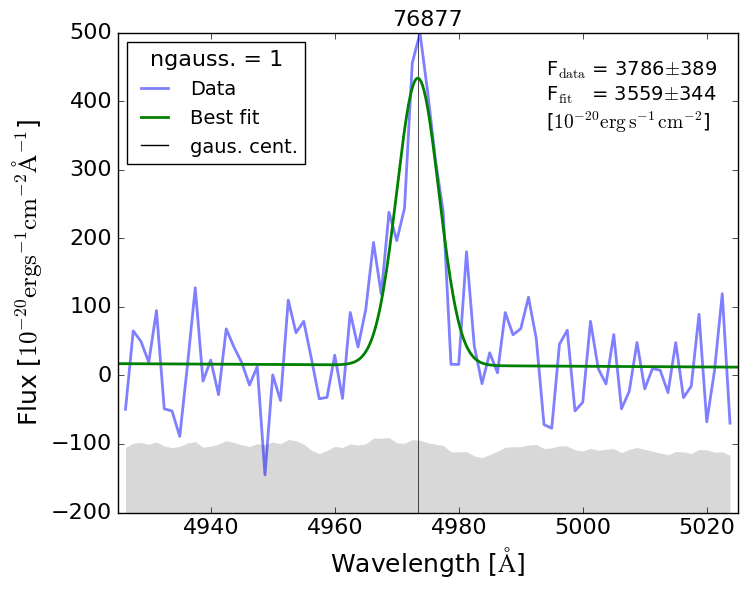}
\includegraphics[scale=0.18]{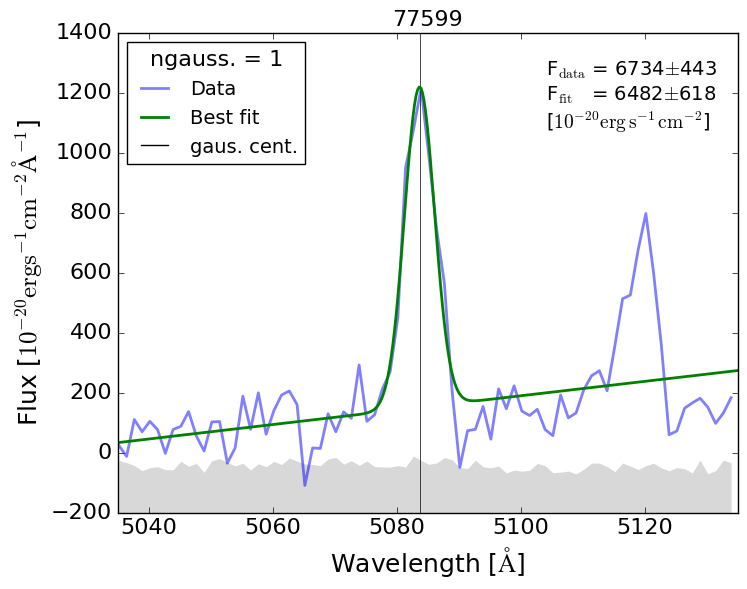}
\includegraphics[scale=0.18]{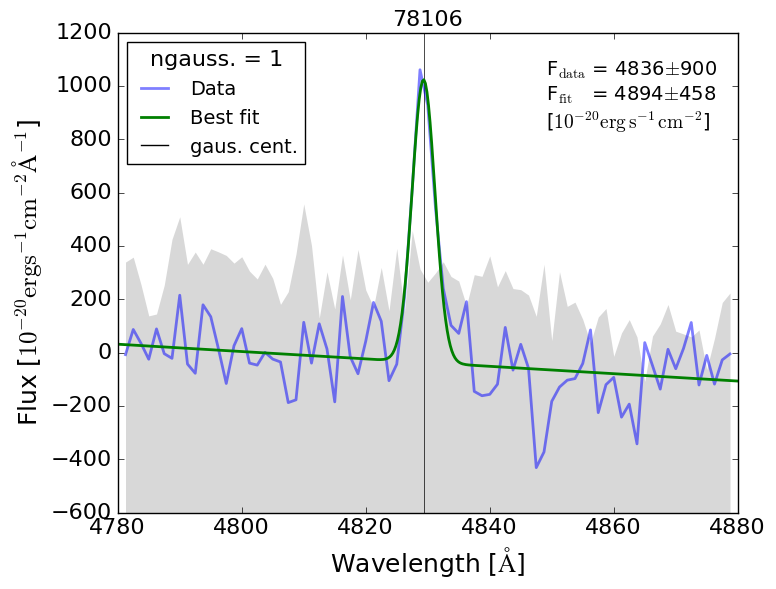}
\includegraphics[scale=0.18]{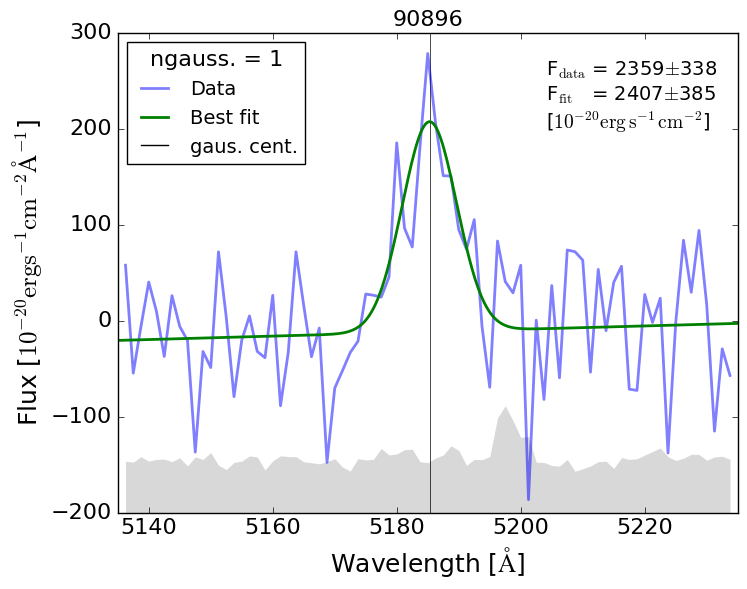}
\includegraphics[scale=0.18]{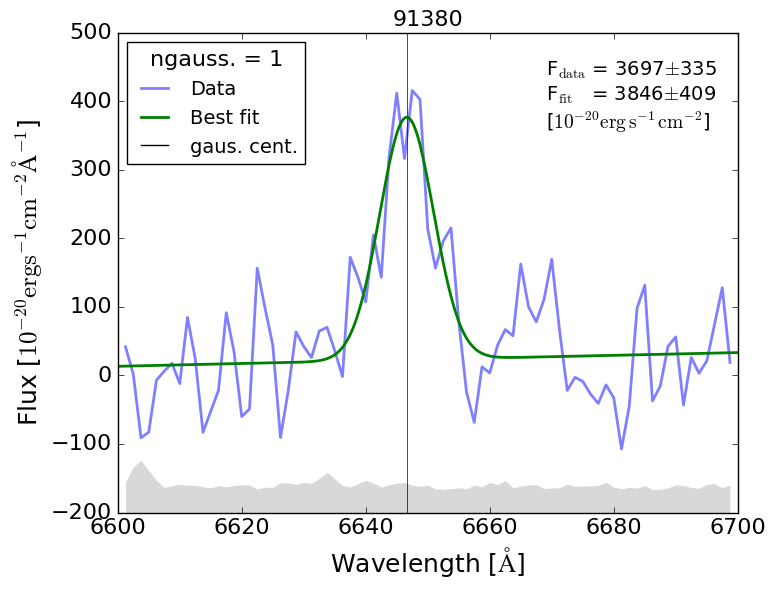}
\includegraphics[scale=0.18]{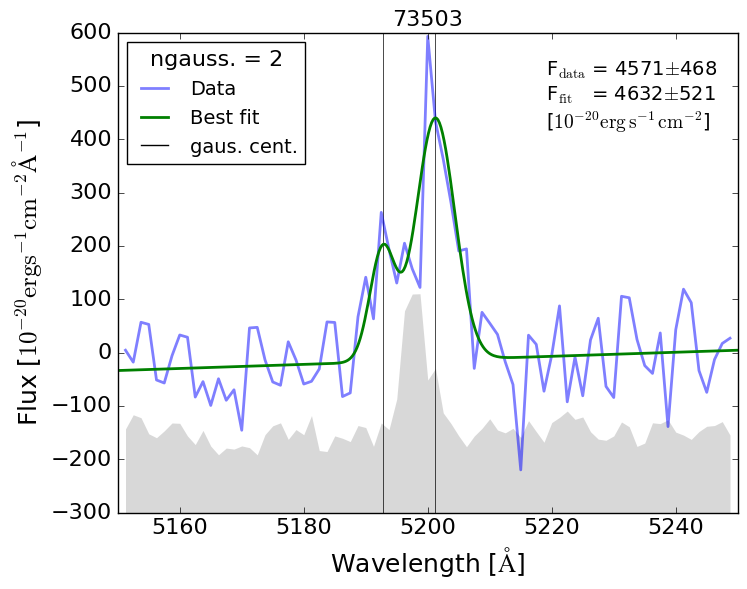}
\includegraphics[scale=0.18]{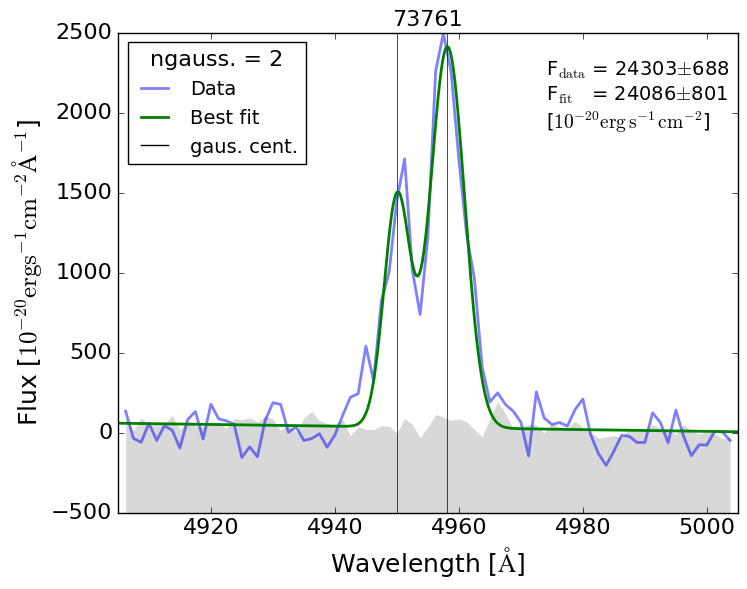}
\includegraphics[scale=0.18]{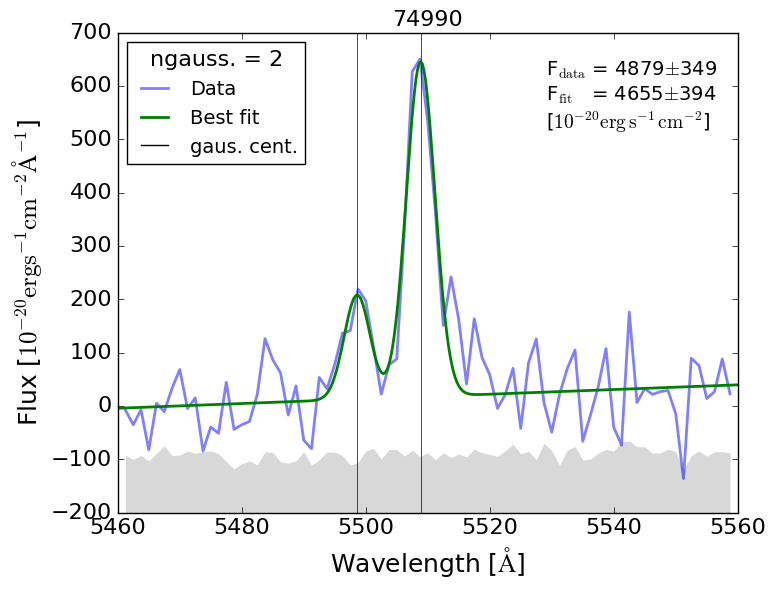}
\includegraphics[scale=0.18]{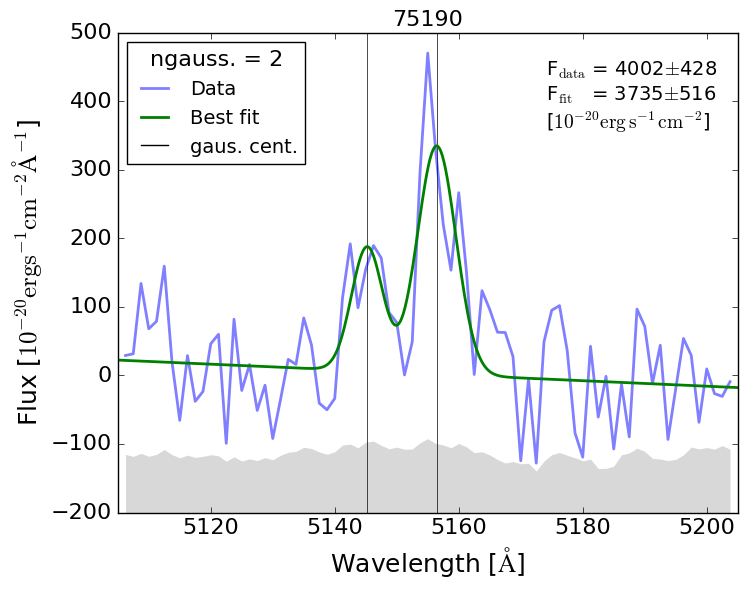}
\includegraphics[scale=0.18]{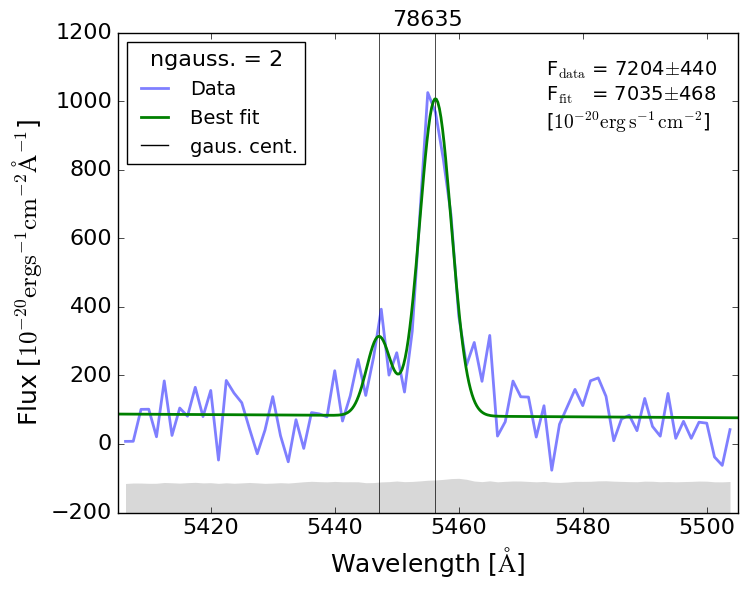}
\includegraphics[scale=0.18]{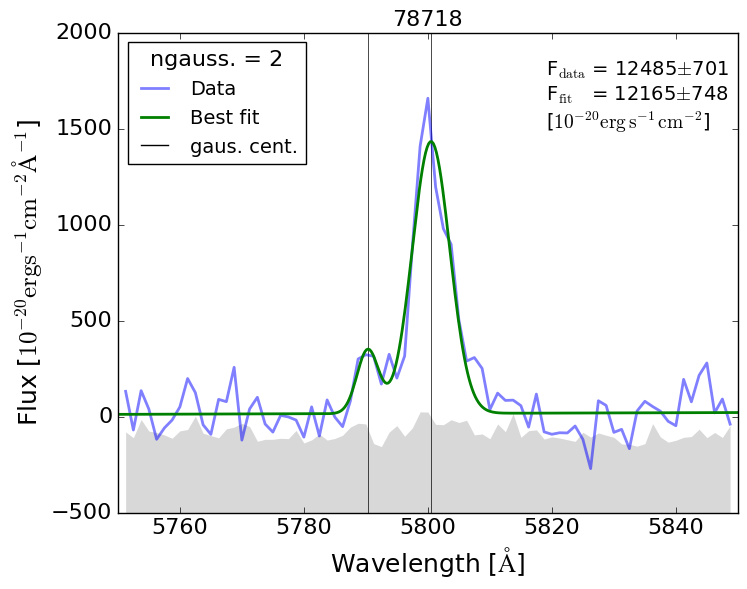}
\includegraphics[scale=0.18]{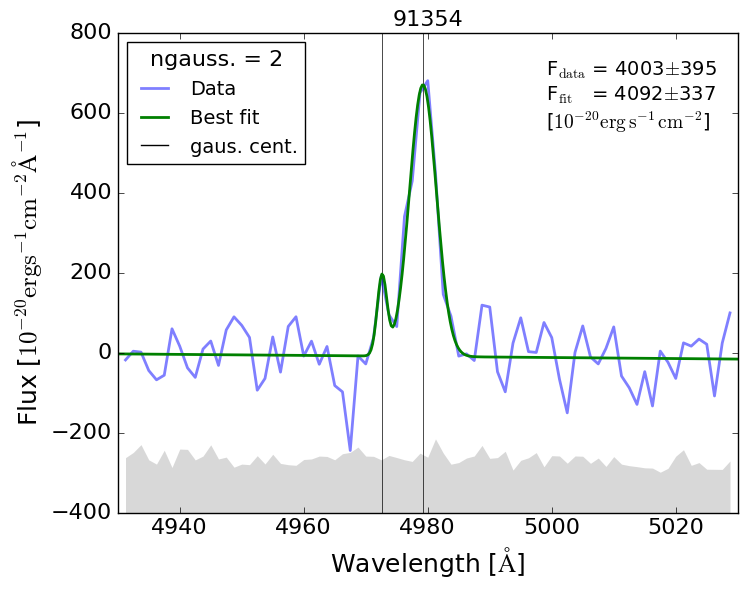}
\includegraphics[scale=0.18]{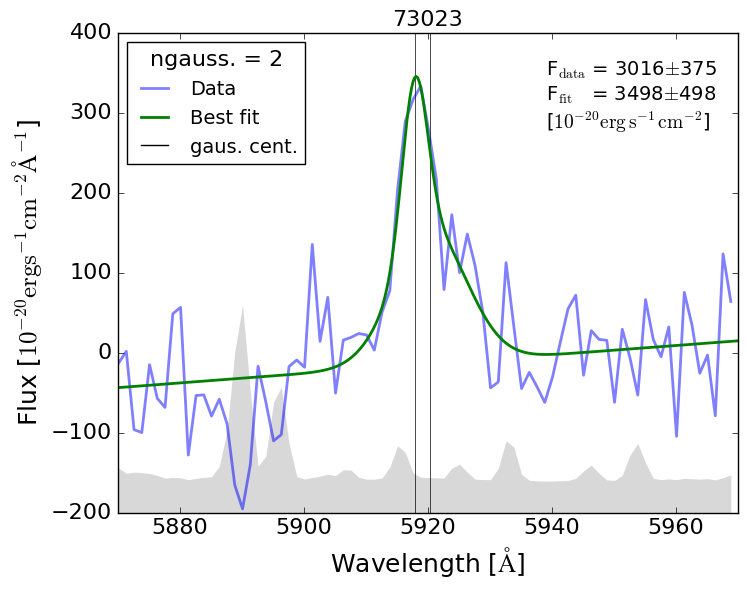}
\includegraphics[scale=0.18]{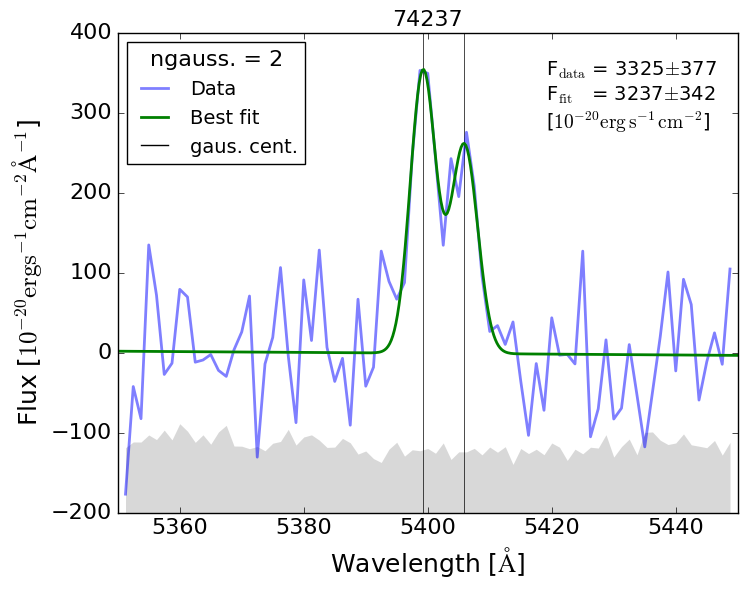}
\includegraphics[scale=0.18]{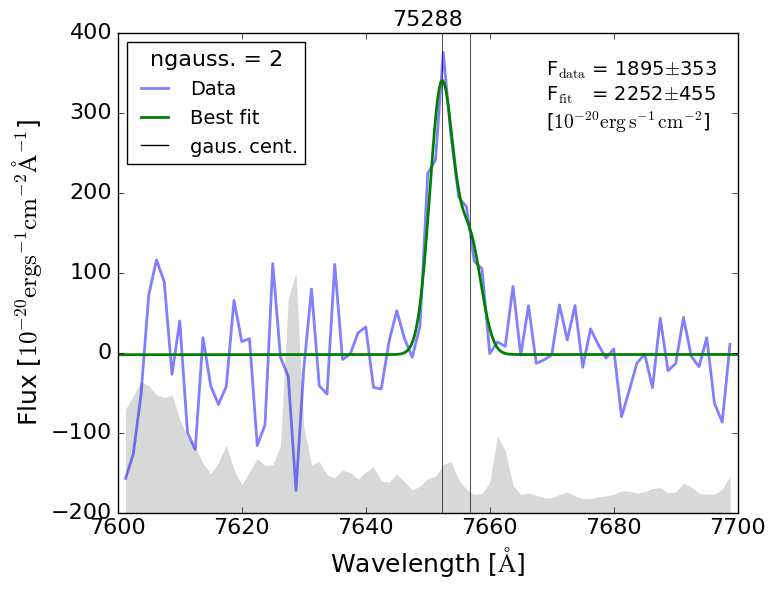}
\includegraphics[scale=0.18]{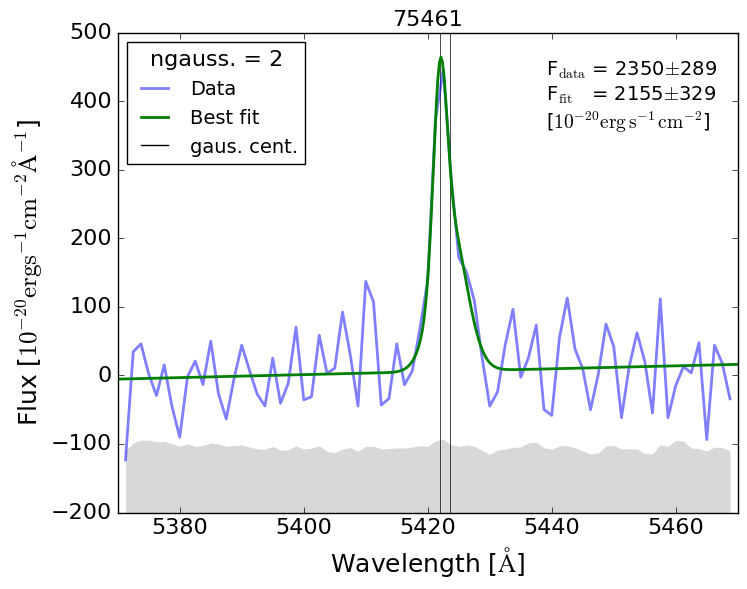}
\includegraphics[scale=0.18]{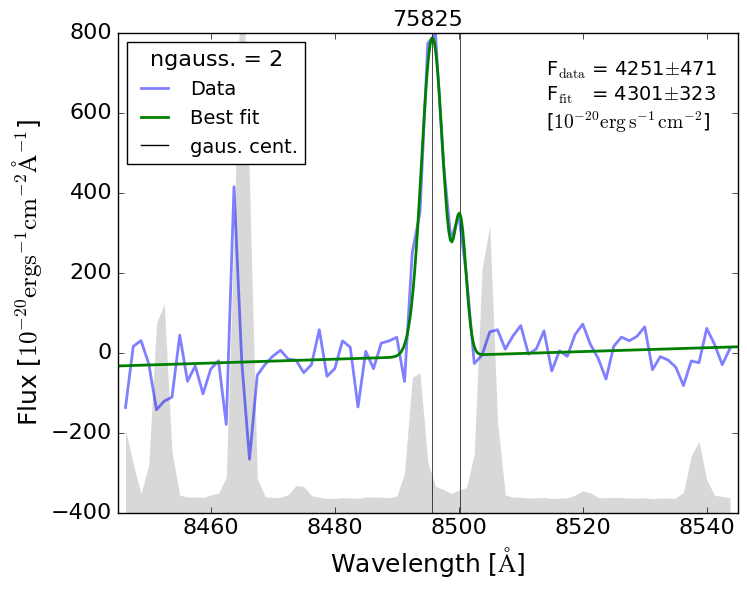}
\includegraphics[scale=0.18]{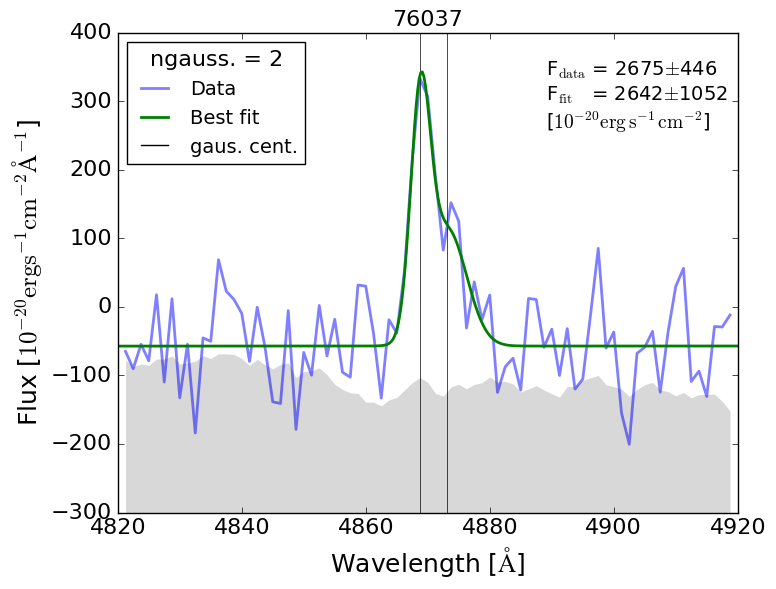}
\includegraphics[scale=0.18]{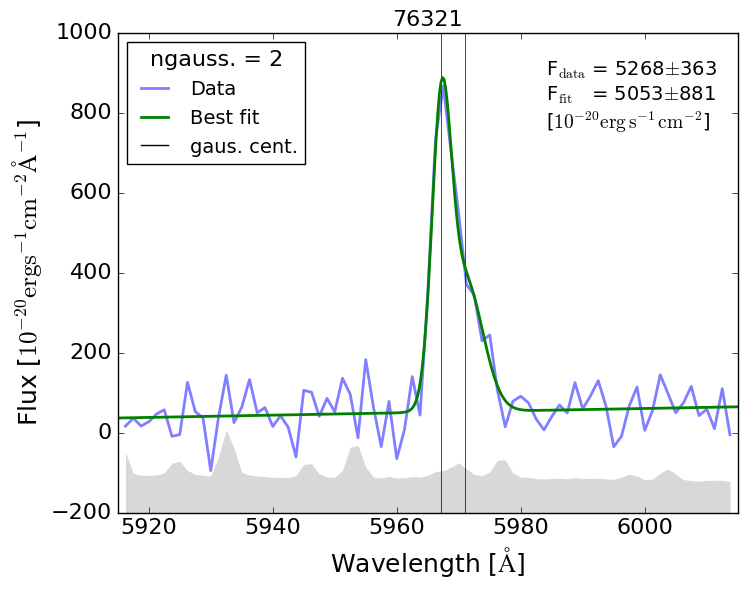}
\includegraphics[scale=0.18]{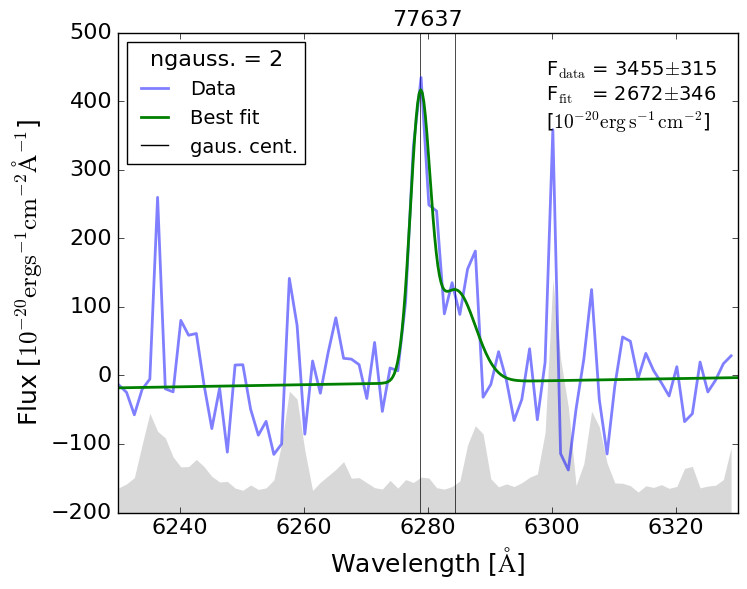}
\includegraphics[scale=0.18]{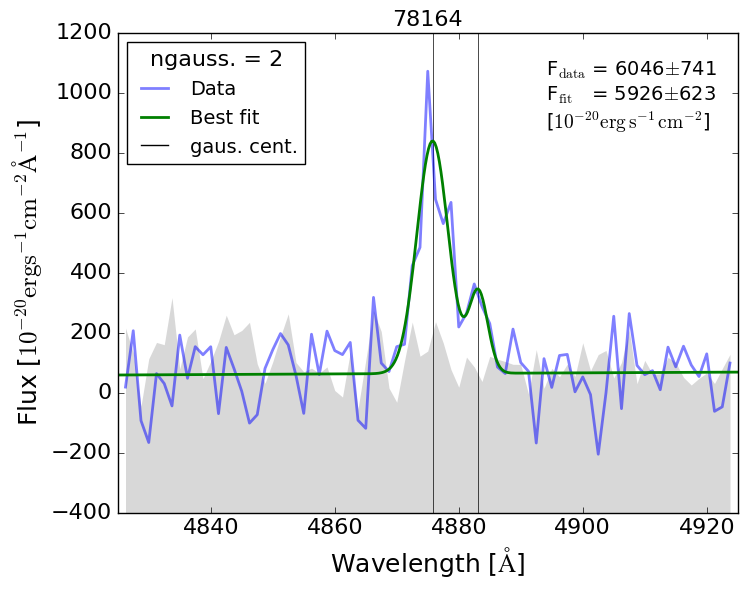}
\includegraphics[scale=0.18]{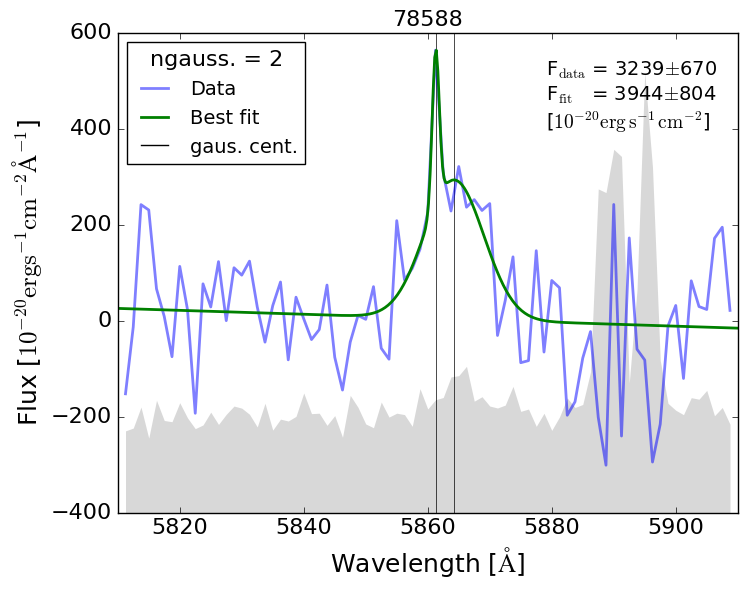}
\includegraphics[scale=0.18]{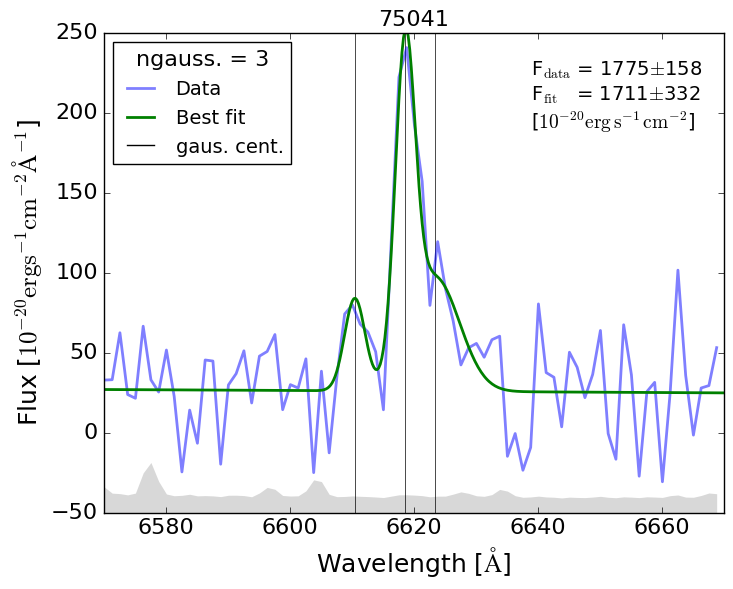}
\includegraphics[scale=0.18]{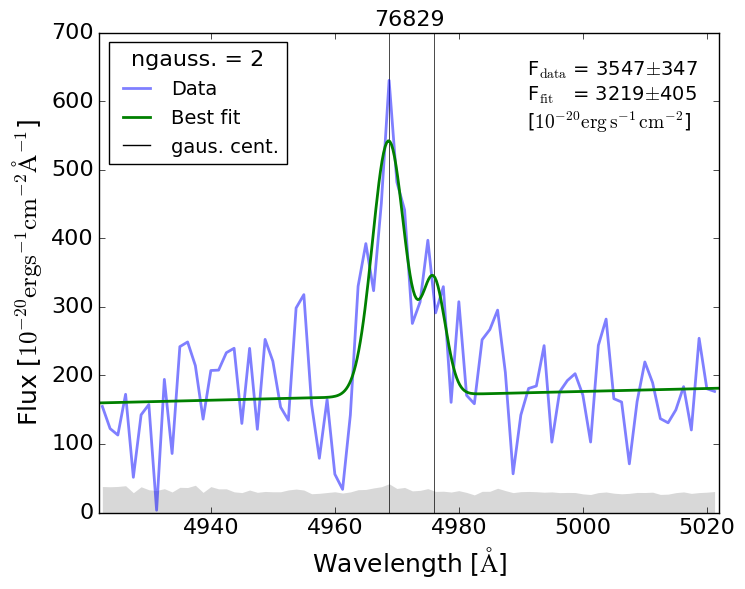}
\caption{Zoom-in of the spectra of our MUSE/SMUVS Ly$\rm \alpha$ emitters. The wavelength range is chosen so that the Ly$\rm \alpha$ line is visible and centered in the plot. The fit is performed by extracting the spectrum from the MUSE datacube in the region where the Ly$\rm \alpha$ emission is $\rm \sim 3\sigma$ above the background or covers at least 50 pixels. The number of gaussians used in the fit is determined after visual inspection of the shape of the spectrum and serves purely to measure the flux. All three line profiles are represented here.}
\label{fig:LyAsnip1}
\end{figure*}
We focused our attention on the spectra of the 39 galaxies which have $z_{\rm spec} \ge 2.0$. We separate the Ly$\rm \alpha$ emitters from absorption line galaxies and AGNs, ending up with 36 sources at $\rm z \gtrsim 3$ and 3 sources at redshift $\rm 2\le z\le 3$, as described above.

\subsubsection{Lyman~$\rm \alpha$ emitters}
We extracted the spectra of our Ly$\rm \alpha$ emitters by identifying the extended area of the line emission. This area is defined by the pixels that have signal-to-noise ratio higher than three. In order to account for the instrument PSF, we also required a minimum area of 50 pixels. We then added the flux from the single pixels in the area of emission. Finally, we fitted one or two gaussians to the obtained spectrum and measured the observed flux of the line.\par
Table~\ref{table:FluxLyA} contains our line flux measurements, along with the luminosity of the Ly$\rm \alpha$ line derived using the flux and the redshift measured for our objects. We measure the line flux by fitting our data with a number of gaussians depending on the line profile shown in the spectrum. We then integrate the gaussians in the wavelength interval containing the line to get the value of the flux reported in Table~\ref{table:FluxLyA}. The mean signal to noise of our lines is $\rm \sim 9.3$, with values spanning from 2.5 to 30. The luminosities we measure are of the order of $\rm 10^{42}-10^{43}$ erg s$\rm ^{-1}$, in line with the values published for MUSE-Wide data in other fields \citep{2017A&A...606A..12H,2019A&A...621A.107H}.\par
Finally, we note that three different line profiles can be identified in our sample: a single line profile, where the emission appears mostly symmetric, a blue bump profile, where in addition to the main, more intense line a secondary peak in the blue is visible and a red tail profile, where the emission is either asymmetric with an extended tail in the red part of the spectrum or presents a secondary peak in the red. Fig.~\ref{fig:LyAsnip1} shows the zoomed-in region of the spectra in the Ly$\rm \alpha$ wavelength range. The plots in the figure further show that we cannot detect a continuum level in our data, thus not allowing us to recognize P-Cygni profiles, even if they are commonly observed for low redshift objects.\par
For the three profiles that we recognize, we report the following statistics. Out of 36 objects, 17 show a single line profile ($\rm \sim 47$\%), 7 show a blue bump ($\rm \sim 19$\%), 11 show a red tail ($\rm \sim 31$\%) and finally one shows both a blue bump and a red tail ($\rm \sim 3$\%).\par
The different profiles are caused by the condition and state of the medium in and around the galaxy. For example, a narrow single line can be caused by a reduced amount of scattering for the Ly$\rm \alpha$ photons due to a low hydrogen column density. The blue bump can be caused either by re-emission by the medium of blueshifted Ly$\rm \alpha$ photons or by a strong absorption at the resonant wavelength, leaving only the red tail and a fraction of the original emission. Finally, the red tail can give us information on the offset velocity of the medium, its optical depth and the hydrogen column density. Radiative transfer models like the ones by \citet{2008A&A...491...89V} or the more recent ones by \citet{2017A&A...608A.139G}, can help disentangle the state of the gas and dust around the emitting galaxy. Applying such models is, however, outside of the scope of this paper, mainly due to the high noise in our spectra, and is left for follow-up studies.

\subsubsection{SMUVS/MUSE sources at $\rm 2\le z \le 3$}

Additionally to the 36 Ly$\rm \alpha$ emitters, we identified 3 other high redshift objects in the range $\rm 2\le z \le 3$. Object \#75267 is an absorption line galaxy at redshift $\rm z=2.4796$, classified as a $\rm QF=2$ spectrum. We report seeing the CIV doublet in absorption. Object \#78359 is classified as a $\rm QF=3$ AGN at redshift $\rm z=2.1458$ with broad line emission of both [CIII] and CIV indicating the AGN activity. And finally, object \#78448 is a galaxy at redshift $\rm z=2.1723$ with $\rm QF=2$ emitting CIII and with possible FeII absorptions. 

\subsection{Physical properties of the SMUVS MUSE galaxies inferred from broad-band photometry}
\label{sec:inSMUVS}

We investigate the distribution of properties derived from SED fitting for our sample. Since the physical properties of each object are derived from SMUVS photometry by fitting its SED, we first need to check whether the redshift measured with MUSE matches the photometric redshifts originally derived by \citet{2018ApJ...864..166D}.\par
We find that, out of the 691 SMUVS/MUSE sources for which we measured spectroscopic redshifts with high confidence, 624 sources have a redshift compatible with the photometric value, while 62 are outliers. To define outliers we follow the same definition used by \citet{2018ApJ...864..166D}, i.e.,
\begin{equation}
\rm \sigma_z=\frac{|z_{spec}-z_{phot}|}{(1+z_{spec})} > 0.15,
\label{eq:sigz}
\end{equation}
where $\rm z_{spec}$ is our spectroscopic redshift, while $\rm z_{phot}$ is the photometric redshift of the SMUVS catalog.\par
Here we find a percentage of outliers of $\rm \sim 10\%$, which is somewhat larger than what was found in \citet{2018ApJ...864..166D} when comparing all their photometric redshifts with the available COSMOS spectroscopic redshifts, over the whole SMUVS/COSMOS area (the outlier fraction there was $\rm 5.5\%$). This difference is perhaps not surprising, given that there is no source pre-selection in MUSE, while spectra taken with all other spectrographs are preferentially available for bright sources (for which the photometry has a higher signal-to-noise ratio and, thus, the photometric analysis is more likely to yield good redshifts).  In any case, it is reassuring that the percentage of redshift outliers that we obtain here is still reasonably low.\par
Before considering the SED fitting properties of our galaxies, we investigated the reason of the redshift discrepancies among the outliers.  We focused on the $z\geq2$ sources, which are the main interest here. Among the 39 high-redshift sources, we found 24 for which the photometric and spectroscopic redshifts are in good agreement. We analyzed the remaining 15 cases on an individual basis, in order to understand whether there is any problem in the spectroscopic and/or photometric analysis, or whether the SMUVS/MUSE sources matching is correct. For 8 of the outliers, we found that the photometry is contaminated by a brighter neighbor. For the other 7 outliers, we found no apparent photometric contamination. The redshift discrepancy is produced by either the galaxy being fainter than the limiting magnitude of the survey (4 cases) or the SMUVS detection actually being two unresolved, separate sources (3 cases). In the first situation the photometry of the source is likely unreliable in some bands and in the second case both sources influence the values measured in the photometry, but only one of them emits Ly$\rm \alpha$ and is detected in MUSE. In both cases the photometric fit results in lower $\rm z_{phot}$ when comparing to the $\rm z_{spec}$.\par
As a next step, we redid the SED fitting of our high-redshift galaxies with uncontaminated photometry, fixing their redshifts to the MUSE-based spectroscopic value. To do this, we used the code LePhare\footnote{\it http://www.cfht.hawaii.edu/~arnouts/lephare.html} \citep{1999MNRAS.310..540A,2006A&A...457..841I}, with the same template family and parameter values as in \citet{2018ApJ...864..166D}, and considered the same SMUVS 28-band input catalog. We then rerun LePhare with fixed redshifts also on the sources that didn't show a severe contamination and 6 of them could be recovered. We thus get a final sample of 30 sources (instead of the previous 24) with physical properties derived from photometry, 28 of which are MUSE Ly$\rm \alpha$ emitters.\par
We see that the SMUVS/MUSE sources have extinction values between $\rm 0.0\le\, E(B-V)\le\, 0.3$, with $\rm \sim 70$\% having $\rm E(B-V)\le\, 0.1$, as we would expect from systems that show prominent Ly$\rm \alpha$ lines. Nonetheless, there are still sources that have higher extinctions ($\rm 0.2 \le$ E(B-V) $\rm \le 0.3$), hinting at an even higher unattenuated Ly$\rm \alpha$ luminosity in those cases. The stellar mass range covered by our sources is from $\rm \sim 1.5\times 10^{8}\, M_\odot$ to $\rm \sim 7\times 10^{11}\, M_\odot$, with a mean value of $\rm \sim 10^9\, M_\odot$. Finally, we see our objects have ages ranging from 10 Myrs to 2 Gyrs.\par
In the next subsection, we compare the distribution of the derived SED properties for the Ly$\rm \alpha$ emitters and other SMUVS sources with $\rm z\ge 2.9$ (see also Fig.~\ref{fig:HZhisto}).

\subsection{Comparison of SED properties for SMUVS galaxies with and without MUSE identification}
\label{sec:SED}
\begin{figure}
\centering
\includegraphics[scale=0.45]{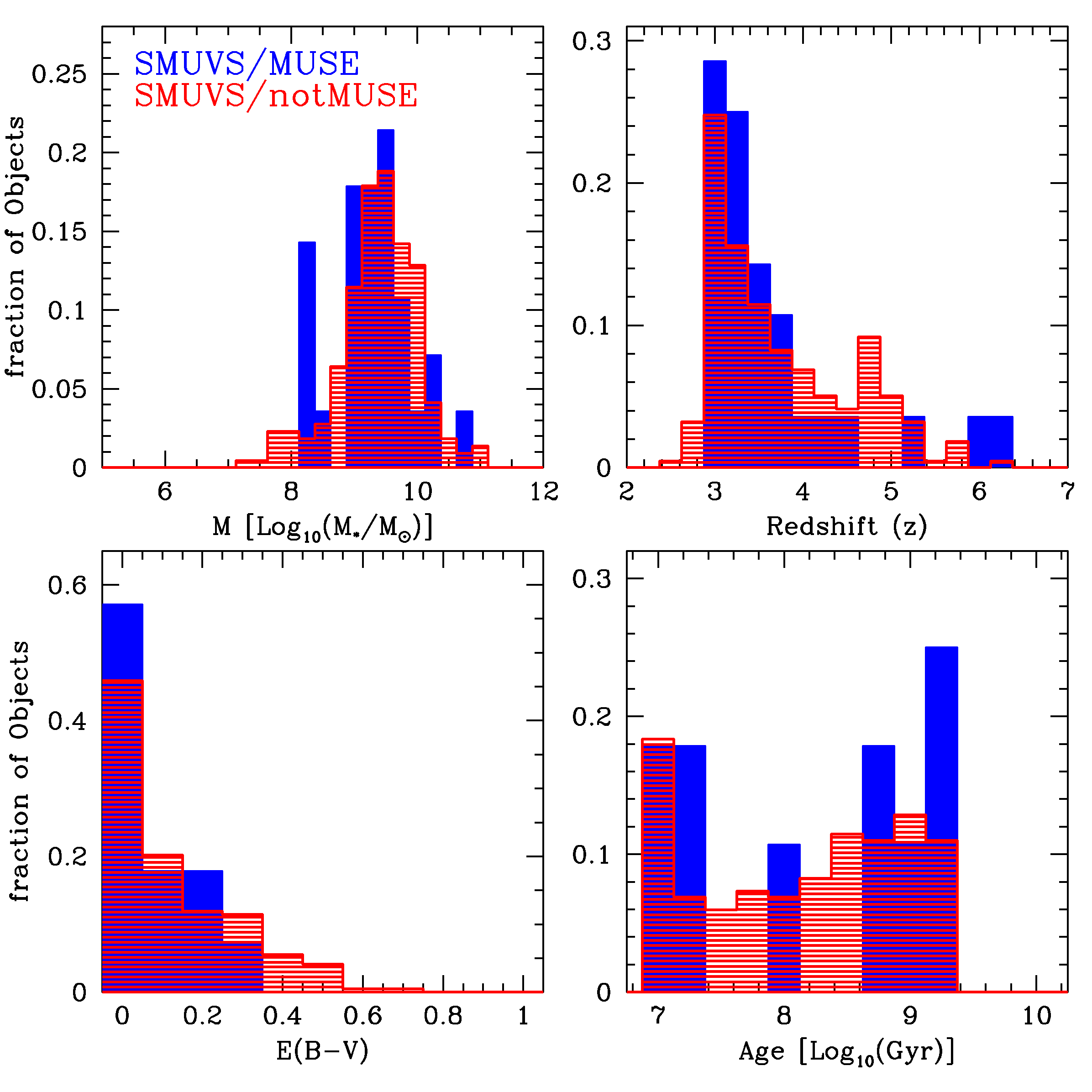}
\caption{Histogram of the three main properties (age, extinction and stellar mass) derived from the SMUVS photometry and the redshifts measured by MUSE. We compare our sample of 28 Ly$\rm \alpha$ emitters to the SMUVS sample of complementary high-z objects (218 galaxies).}
\label{fig:HZhisto}
\end{figure}
\begin{figure}
\centering
\includegraphics[scale=0.45]{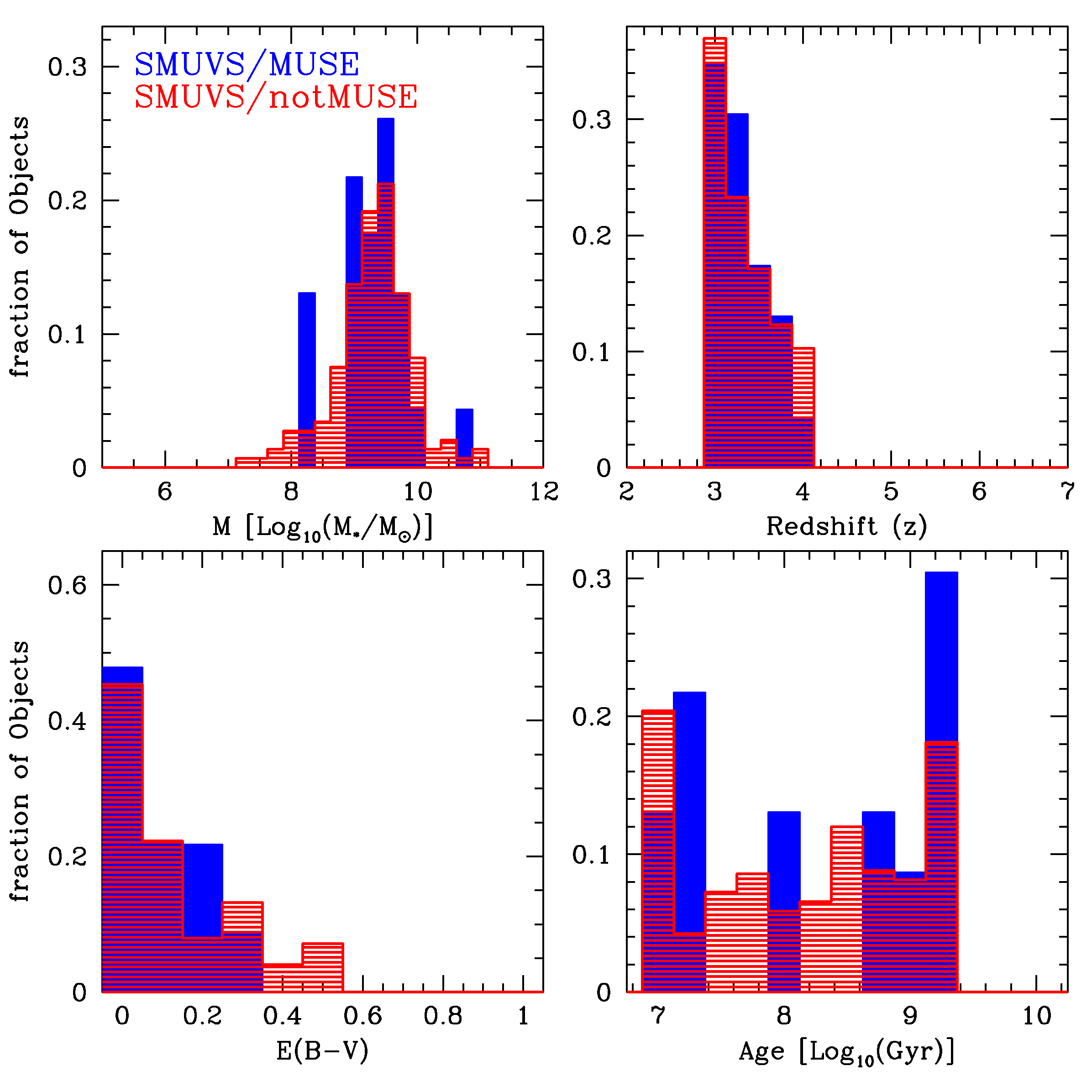}
\caption{The redshift cut and the new stellar mass distribution due to the cut are shown as they were before the mass-matching in the upper panels. The mass matched histograms of the extinction E(B-V) and the age derived from the SMUVS photometry for 23 SMUVS/MUSE and 146 SMUVS/notMUSE sources in the redshift range $\rm 3 \le\, z\le\, 4$ are shown in the lower panels.}
\label{fig:weighthisto}
\end{figure}
\begin{table}
\caption{Table showing the results of all the KS tests performed in this work.}
\label{table:KS}
\centering
\tiny
\begin{tabular}{|c|c|c|c|}
\hline\hline
&\multicolumn{3}{|c|}{p-values for}\\
Test&\multicolumn{2}{c|}{SMUVS/...}&SMUVS/MUSE\\
performed on&\multicolumn{2}{c|}{MUSE vs notMUSE}& vs \\
&original&mass matched&MUSE/NS\\
\hline
Age&0.43&3.5 $\times$ 10$^{-2}$&-\\
\hline
Extinction&0.61&2.3 $\times$ 10$^{-6}$&-\\
\hline
Stellar Mass&0.08&0.997&-\\
\hline
Redshift&0.12&-&0.02\\
\hline
$\rm L_{Ly\alpha}$&-&-&1.5 $\times$ 10$^{-3}$\\
\hline
$\rm F_{Ly\alpha}$&-&-&4 $\times$ 10$^{-6}$\\
\hline
\end{tabular}
\end{table}
We compare the derived SED properties of the SMUVS $\rm z>2.9$ sources which are identified with MUSE (i.e. those with a spectroscopic redshift measurement with QF$\geq2$), with those of the SMUVS $\rm z\gtrsim 2.9$ galaxies which are not MUSE-detected,  in the same field. Our aim is to investigate whether there are significant differences in these properties, particularly to understand whether the most prominent Ly$\rm \alpha$ emitters at $\rm z>3$ are characterized by special values in their physical properties (stellar mass, star-formation histories, dust extinction, etc.).\par
Prior to comparing the SED properties of our objects, we tested whether SMUVS/MUSE and SMUVS/notMUSE galaxies come from the same parent absolute-magnitude distribution. The sample of galaxies we consider is 218 SMUVS galaxies with redshift $2.9\lesssim z_{phot}\lesssim 7$ and the 28 MUSE Ly$\rm \alpha$ emitters for which our spectroscopic redshifts and the SMUVS photometric redshift estimates are in good agreement (as defined in Section \ref{sec:inSMUVS}). The comparison is performed on all 28 photometric bands available to SMUVS separately. We performed a Kolmogorov-Smirnov (KS) test on the statistics of the absolute magnitudes derived with LePhare for our two samples. The results of these tests are that the D parameter assumes values $\rm \lesssim 0.25$ over all the bands tested, the p-value is always $\rm \ge 0.08$. This indicates that our SMUVS high-redshift MUSE-detected and non-detected sources show no statistically significant difference.\par
To test if the sample size can influence the outcome of the test, we reduced the SMUVS sample randomly 10,000 times to subsamples containing on average 40 objects. The KS test is then performed on these smaller samples and the results are compared to the test on the whole SMUVS sample by means of their statistics.  We conclude that sample size does not matter statistically in our case.\par
Fig.~\ref{fig:HZhisto} shows the distribution of the physical properties of our two samples derived from the SED fit. As can be expected, the redshift distribution of both samples is biased towards lower redshifts in the range $\rm 2.9 \le z < 6.6$ covered by MUSE. The distribution of the stellar mass for our SMUVS/MUSE sample is between $\rm 8 < log_{10}(M_*) < 11$, with a peak at $\rm \sim 10^{9.5} M_*$. This sample is also mostly composed of galaxies with little dust content. Nevertheless, some of our objects have $\rm E(B-V)> 0.25$ and are still visible as Ly$\rm \alpha$ emitters. Unless the emission is seen through a gap in the dust distribution, this would make them extremely bright in Ly$\rm \alpha$ to overcome such higher values of dust extinction and still be visible. Finally, the age distribution exhibits a bimodality, either classifying our galaxies as very young ($\rm \lesssim 100$ Myr) or $\rm \sim 1$ Gyr old. We will further comment on this feature in the next subsection. Fig.~\ref{fig:HZhisto} confirms our previous analysis on the input photometric bands used in SMUVS: The distribution of the properties of the SMUVS/MUSE and SMUVS/notMUSE sample are not statistically different.\par
After we performed a KS test on the physical properties, we conclude that, compared to the SMUVS/notMUSE sample, the spectroscopically detected galaxies have generally lower dust extinction, about the same mass and age distributions, and less objects are detected in higher redshift regimes ($\rm 4\le z\le 6.6$), but we see no significant statistical difference in any of the properties. The individual results of our KS tests can be seen in Table~\ref{table:KS}.\par
 To further test the distribution of the physical properties in both samples, we constructed Fig.~\ref{fig:weighthisto}. We restricted ourselves to analyzing galaxies in the redshift range $\rm \sim (3,4)$, to limit the effect of galaxy evolution with redshift and because most of the SMUVS/MUSE objects lie in that range. We then also stellar-mass matched the SMUVS/notMUSE galaxies to SMUVS/MUSE. The redshift cut applies to all the panels, while the mass matching is shown only for the extinction and the ages. We can see how the age distribution appears very similar in both samples and how the distribution of redshift and masses do not deviate much from what we saw in Fig.~\ref{fig:HZhisto}, before the redshift cut was applied. The interesting panel is the one showing the extinction. We can see now how applying a cut in redshift and stellar-mass matching the SMUVS/notMUSE sample reveals a slight difference in the distribution. SMUVS/MUSE shows preferentially lower extinctions compared to SMUVS/notMUSE, who deviates from E(B-V)$\rm > 0.2$ onward and drives the difference. Again, the results of a formal KS test can be seen in Table~\ref{table:KS}.

\subsubsection{Further study of the age bimodality}
\begin{figure}
\centering
\includegraphics[scale=0.48]{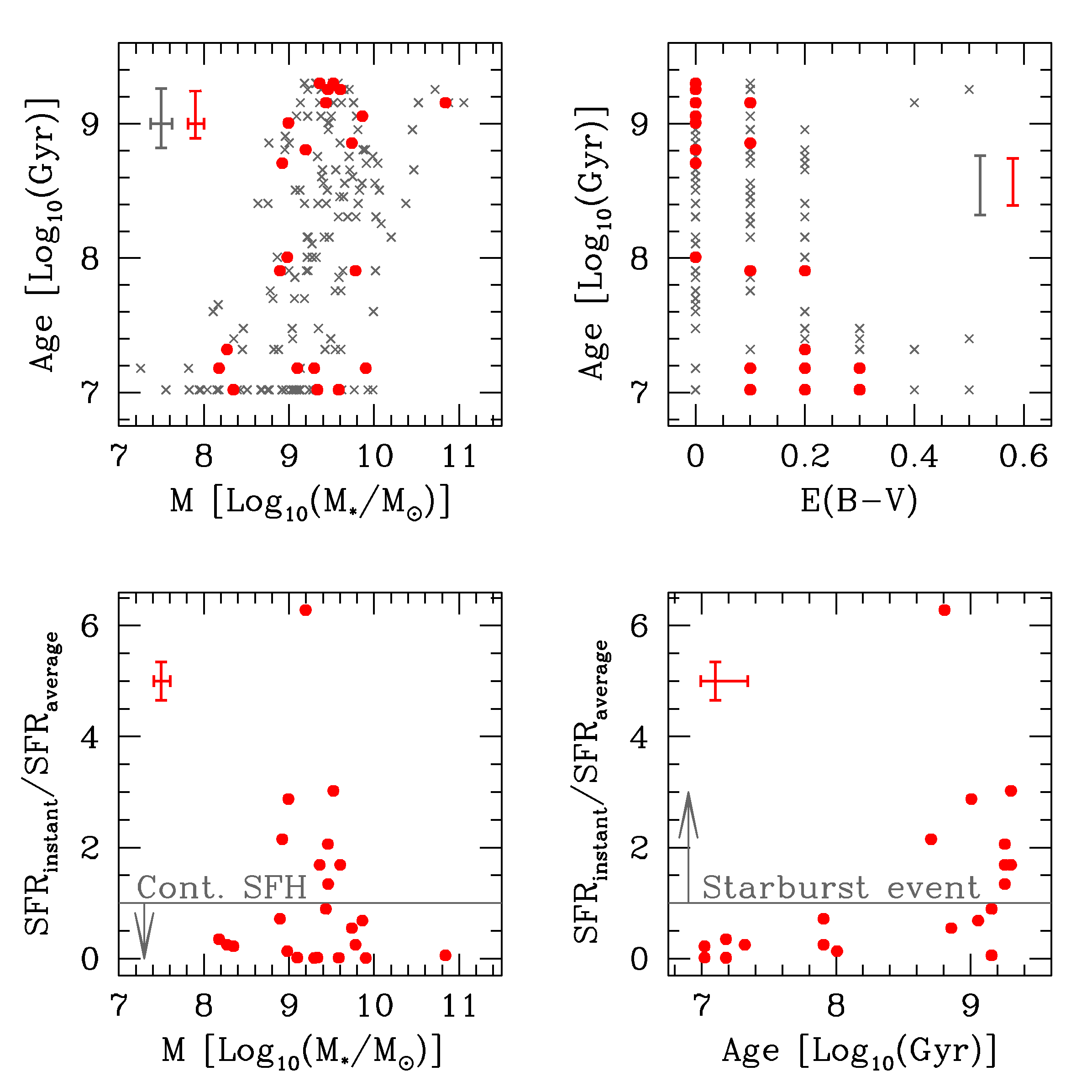}
\caption{{\it Upper panels:} Mass-age relation on the left and E(B-V)-age relation on the right for both our SMUVS detected samples. Here, SMUVS/MUSE is represented in red, while SMUVS/notMUSE is plotted in gray. We see that the older objects are generally more massive and that the younger objects experience generally more extinction than older objects. {\it Lower panels:} Test of the lower limit of the instantaneous SFR derived from our Ly$\rm \alpha$ flux measurements against an average star formation rate obtained by dividing the stellar mass by the age of the galaxy obtained from the SED fit. We see on the left that the higher SFR is associated to intermediate-mass objects and on the right that older objects have a higher SFR than what would be expected if they had continuously formed stars in one single episode. The gray line in the plots indicates where the ratio between SFRs is unity. The errorbars shown in the plots are representative of the average errors on the galactic properties.}
\label{fig:SFR}
\end{figure}
We further study the implications of an age bimodality in our MUSE-detected SMUVS sample by comparing the stellar mass-age and E(B-V)-age relations of SMUVS/MUSE to the relations found in SMUVS/notMUSE. Fig.~\ref{fig:SFR} shows what we find for SMUVS/MUSE in red and SMUVS/notMUSE in gray. We can see that, as is expected from the current view on galaxy evolution, the younger objects tend to be less massive than the older objects and that they are in general more obscured than older objects. We associate these characteristics with the fact that the younger galaxies in our sample are currently experiencing their first star formation event. The older objects in contrast have already formed the bulk of their stars, thus show older populations and less dust is present in their surroundings.\par
We notice however that all SMUVS/MUSE objects are detected in Ly$\rm \alpha$ emission and so we expect them to be actively star-forming. We argue that the star formation that we see in the older objects is not part of the first event, but a separated second episode of star formation. To test this, we generated the lower plots in Fig.~\ref{fig:SFR}. We compare the SFR obtained if we assume a continuous star formation throughout the lifetime of our objects and the lower limit of the SFR given by the Ly$\rm \alpha$ emission we detect. The first value is obtained by dividing the stellar mass given by the SED fit with the age of the galaxy and represents the average of the SFR. The second value is obtained by applying the conversion $\rm L_{H\rm \alpha}=L_{Ly\rm \alpha}*\frac{1}{8.7}$, which gives us a lower limit for the luminosity in the H$\rm \alpha$ line and for which we chose to use the Ly$\rm \alpha$-H$\rm \alpha$ conversion factor by \citet{1998ApJ...502L..99H}. We then use the \citet{1998ARA&A..36..189K} equation to obtain the SFR from the H$\rm \alpha$ luminosity to obtain the lower limit of the instantaneous SFR.\par
By comparing these two values we can qualitatively say if a galaxy is consistent with a monotonically declining star formation history ($\rm SFR_{instant}/SFR_{average}\le 1$), or if a second star formation episode is needed to explain their SFR ($\rm SFR_{instant}/SFR_{average}> 1$). We see that galaxies that are experiencing a second event are intermediate-mass objects ($\rm 9 \lesssim log_{10}(M_*) \le 10$) and are exclusively old. We argue that the main component of the stellar population has formed in early times with the first star formation event and is what the SMUVS photometry detects, while the new stars being formed are detected by MUSE. We conclude that these objects are probably experiencing a rejuvenation event and that their star formation has been restarted after the stellar bulk has been formed about 1 Gyr before the time we observe them.

\subsection{Additional MUSE high redshift sources not present in SMUVS}
\label{sec:outSMUVS}
\begin{table*}
\caption{Blind search detected high-redshift ($\rm z \ge 2.9$) Ly$\rm \alpha$ emitting sources in the COSMOS/MUSE GTO field. As in Table~\ref{table:HZsource}, the ID, position on the sky, measured redshift and quality flag are displayed. Additionally, also the information on the Ly$\rm \alpha$ flux and luminosity measurement is listed. The labels on the line profile refer to the single peak (SP), blue bump (BB) or red tail (RT).}
\label{table:noSMUVS}
\centering
\tiny
\begin{tabular}{c|c c|c|c||c|c|c|c|c|}
\hline\hline                 
ObjID & RA & Dec & $\rm z_{spec}$ & QF & Line Center (Main) & Line Center (Sec.) & Line Profile & $\rm F_{fit}$ & $\rm L_{fit}$ \\
\#&&&&&$\rm [\AA]$&$\rm [\AA]$&[SP,BB,RT]&$\rm [10^{-18}\, erg\, s^{-1}\, cm^{-2}]$&$\rm [10^{42}\, erg\, s^{-1}]$\\
\hline
  NS1 & 150.0956195 & 2.207896860 & 3.7137 & 2&5730.5 $\pm$ 0.3&-&SP&18.96 $\pm$ 1.83&2.55 $\pm$ 0.25\\
  NS3 & 150.0890289 & 2.202755060 & 4.0670 & 9&6160.0 $\pm$ 0.4&-&SP&13.39 $\pm$ 1.85&2.23 $\pm$ 0.31\\
  NS4 & 150.0902794 & 2.208985301 & 4.0681 & 3&6160.8 $\pm$ 0.3&6163.4 $\pm$ 0.9&RT&22.03 $\pm$ 2.25&3.67 $\pm$ 0.38\\
  NS6 & 150.1064723 & 2.205999095 & 3.1468 & 9&5042.2 $\pm$ 0.3&5035.3 $\pm$ 0.5&BB&14.57 $\pm$ 1.86&1.32 $\pm$ 0.17\\
  NS7 & 150.1062366 & 2.202861769 & 3.6407 & 3&5641.9 $\pm$ 0.4&5646.8 $\pm$ 2.1&RT&24.50 $\pm$ 5.05&3.13 $\pm$ 0.65\\
  NS8 & 150.1107880 & 2.196746174 & 4.9088 & 9&7182.3 $\pm$ 0.4&-&SP&15.64 $\pm$ 1.65&4.06 $\pm$ 0.43\\
  NS10 & 150.1271781 & 2.193819620 & 4.5575 & 3&6756.1 $\pm$ 0.3&6759.7 $\pm$ 0.9&RT&26.69 $\pm$ 2.32&5.82 $\pm$ 0.51\\
  NS15 & 150.1631312 & 2.200003868 & 3.1316 & 9&5022.0 $\pm$ 0.5&-&SP&13.49 $\pm$ 2.33&1.20 $\pm$ 0.21\\
  NS16 & 150.1611895 & 2.206540221 & 3.6455 & 9&5645.4 $\pm$ 0.3&5648.9 $\pm$ 1.1&RT&22.51 $\pm$ 3.12&2.89 $\pm$ 0.40\\
  NS19 & 150.1766375 & 2.200547905 & 3.1329 & 3&5024.4 $\pm$ 0.4&5026.9 $\pm$ 2.1&RT&37.82 $\pm$ 7.11&3.37 $\pm$ 0.63\\
  NS20 & 150.1804873 & 2.207906097 & 3.9656 & 2&6036.8 $\pm$ 0.2&-&SP&17.42 $\pm$ 1.30&2.74 $\pm$ 0.20\\
  NS21 & 150.1897355 & 2.204086497 & 4.0641 & 3&6156.4 $\pm$ 0.2&-&SP&23.91 $\pm$ 1.86&3.98 $\pm$ 0.31\\
  NS22 & 150.1819584 & 2.196144730 & 4.3955 & 9&6559.7 $\pm$ 0.4&6563.7 $\pm$ 3.8&RT&7.04 $\pm$ 3.56&1.41 $\pm$ 0.71\\
  NS23 & 150.1763508 & 2.194987214 & 4.4405 & 3&6612.7 $\pm$ 0.3&6615.4 $\pm$ 0.6&RT&12.50 $\pm$ 1.14&2.56 $\pm$ 0.23\\
  NS24 & 150.1871556 & 2.205024286 & 5.0682 & 2&7377.7 $\pm$ 0.3&-&SP&8.26 $\pm$ 1.17&2.32 $\pm$ 0.33\\
  NS27 & 150.0900671 & 2.221593035 & 3.2476 & 3&5163.4 $\pm$ 0.5&-&SP&22.50 $\pm$ 2.82&2.19 $\pm$ 0.27\\
  NS28 & 150.0938165 & 2.212022447 & 3.3896 & 3&5336.1 $\pm$ 0.2&5338.3 $\pm$ 1.0&RT&36.35 $\pm$ 4.19&3.92 $\pm$ 0.45\\
  NS29 & 150.0967870 & 2.210240788 & 3.4315 & 9&5387.0 $\pm$ 1.2&5387.4 $\pm$ 0.3&RT&11.21 $\pm$ 2.80&1.24 $\pm$ 0.31\\
  NS30 & 150.0974611 & 2.210715291 & 3.7856 & 3&5818.2 $\pm$ 0.2&5823.5 $\pm$ 0.5&RT&42.11 $\pm$ 3.16&5.92 $\pm$ 0.44\\
  NS32 & 150.0859870 & 2.224398122 & 4.2863 & 3&6426.5 $\pm$ 0.2&-&SP&18.42 $\pm$ 1.40&3.48 $\pm$ 0.26\\
  NS34 & 150.0870797 & 2.216011780 & 4.9009 & 9&7172.5 $\pm$ 0.6&7177.7 $\pm$ 1.3&RT&8.61 $\pm$ 1.52&2.41 $\pm$ 0.43\\
  NS35 & 150.0968042 & 2.219972452 & 4.9027 & 9&7175.7 $\pm$ 0.4&-&SP&15.66 $\pm$ 1.53&4.06 $\pm$ 0.40\\
  NS36 & 150.0850979 & 2.213391448 & 5.0266 & 9&7324.7 $\pm$ 0.3&7327.8 $\pm$ 0.6&RT&18.20 $\pm$ 1.60&5.00 $\pm$ 0.44\\
  NS40 & 150.1027830 & 2.218819424 & 3.6257 & 9&5622.8 $\pm$ 0.2&-&SP&17.26 $\pm$ 2.44&2.19 $\pm$ 0.31\\
  NS41 & 150.1009535 & 2.213916903 & 3.8283 & 9&5870.8 $\pm$ 0.5&-&SP&22.61 $\pm$ 3.47&3.27 $\pm$ 0.50\\
  NS42 & 150.1045699 & 2.221037671 & 5.2998 & 2&7659.2 $\pm$ 0.3&7666.1 $\pm$ 2.8&RT&20.27 $\pm$ 3.15&6.31 $\pm$ 0.98\\
  NS46 & 150.1273378 & 2.214692674 & 3.0943 & 3&4977.4 $\pm$ 0.3&-&SP&33.23 $\pm$ 3.75&2.88 $\pm$ 0.33\\
  NS48 & 150.1282981 & 2.211522251 & 3.3210 & 9&5252.9 $\pm$ 0.4&-&SP&13.47 $\pm$ 2.13&1.38 $\pm$ 0.22\\
  NS49 & 150.1236488 & 2.221830819 & 4.2820 & 9&6420.5 $\pm$ 0.3&6425.5 $\pm$ 2.0&RT&10.05 $\pm$ 1.88&1.89 $\pm$ 0.35\\
  NS53 & 150.1311033 & 2.221893546 & 3.2790 & 9&5202.1 $\pm$ 0.2&-&SP&11.75 $\pm$ 1.37&1.17 $\pm$ 0.14\\
  NS54 & 150.1444920 & 2.210705671 & 4.1055 & 9&6206.9 $\pm$ 0.3&-&SP&10.56 $\pm$ 1.28&1.80 $\pm$ 0.22\\
  NS55 & 150.1331736 & 2.224723824 & 4.5402 & 9&6735.4 $\pm$ 0.3&6738.3 $\pm$ 0.9&RT&7.16 $\pm$ 1.42&1.55 $\pm$ 0.31\\
  NS56 & 150.1355132 & 2.212112631 & 5.4193 & 9&7804.4 $\pm$ 0.2&-&SP&6.95 $\pm$ 0.94&2.28 $\pm$ 0.31\\
  NS58 & 150.1442147 & 2.221428993 & 5.8863 & 9&8372.0 $\pm$ 0.3&-&SP&14.19 $\pm$ 1.30&5.65 $\pm$ 0.52\\
  NS61 & 150.1572317 & 2.209132493 & 3.8550 & 9&5901.6 $\pm$ 0.7&5902.4 $\pm$ 0.8&RT&18.30 $\pm$ 2.67&2.69 $\pm$ 0.39\\
  NS62 & 150.1750090 & 2.212290512 & 2.9695 & 3&4825.8 $\pm$ 0.2&-&SP&14.97 $\pm$ 1.80&1.17 $\pm$ 0.14\\
  NS64 & 150.1644704 & 2.214284122 & 3.0976 & 9&4981.4 $\pm$ 0.4&4985.1 $\pm$ 4.8&RT&14.44 $\pm$ 7.93&1.25 $\pm$ 0.69\\
  NS65 & 150.1617505 & 2.210353565 & 3.4902 & 9&5458.4 $\pm$ 0.5&-&SP&8.10 $\pm$ 1.59&0.94 $\pm$ 0.18\\
  NS67 & 150.1634469 & 2.210140164 & 3.8656 & 3&5914.6 $\pm$ 0.3&5919.6 $\pm$ 0.4&RT&16.96 $\pm$ 1.57&2.51 $\pm$ 0.23\\
  NS68 & 150.1674625 & 2.215432552 & 4.4342 & 9&6606.2 $\pm$ 0.8&-&SP&10.62 $\pm$ 1.67&2.17 $\pm$ 0.34\\
  NS72 & 150.1818099 & 2.215637886 & 5.2455 & 9&7592.6 $\pm$ 0.3&-&SP&25.86 $\pm$ 2.79&7.87 $\pm$ 0.85\\
  NS73 & 150.0847432 & 2.233357253 & 3.2413 & 3&5155.9 $\pm$ 0.2&5157.3 $\pm$ 1.1&RT&54.08 $\pm$ 4.63&5.24 $\pm$ 0.45\\
  NS74 & 150.0865742 & 2.227631189 & 3.4529 & 9&5413.2 $\pm$ 0.2&-&SP&14.25 $\pm$ 1.57&1.61 $\pm$ 0.18\\
  NS75 & 150.0946790 & 2.230024846 & 4.0667 & 3&6160.4 $\pm$ 0.3&-&SP&13.55 $\pm$ 1.50&2.26 $\pm$ 0.25\\
  NS78 & 150.0964429 & 2.229859080 & 5.7850 & 9&8249.5 $\pm$ 0.3&-&SP&8.59 $\pm$ 1.35&3.29 $\pm$ 0.52\\
  NS79 & 150.1131469 & 2.230677797 & 3.2805 & 9&5203.0 $\pm$ 0.5&-&SP&56.12 $\pm$ 6.08&5.60 $\pm$ 0.61\\
  NS80 & 150.1237063 & 2.229857749 & 3.4468 & 9&5406.0 $\pm$ 0.2&-&SP&20.57 $\pm$ 2.49&2.31 $\pm$ 0.28\\
  NS81 & 150.1244038 & 2.231267693 & 3.9679 & 3&6039.8 $\pm$ 0.3&-&SP&24.76 $\pm$ 3.44&3.89 $\pm$ 0.54\\
  NS82 & 150.1390775 & 2.235081324 & 3.1188 & 3&5007.3 $\pm$ 0.2&4996.1 $\pm$ 1.1&BB&75.78 $\pm$ 4.98&6.70 $\pm$ 0.44\\
  NS83 & 150.1438608 & 2.231322324 & 3.2779 & 9&5199.8 $\pm$ 0.3&-&SP&4.65 $\pm$ 1.33&0.46 $\pm$ 0.13\\
  NS85 & 150.1599066 & 2.230670888 & 3.3592 & 9&5299.7 $\pm$ 0.3&5306.4 $\pm$ 0.5&RT&15.30 $\pm$ 1.79&1.61 $\pm$ 0.19\\
  NS86 & 150.1587096 & 2.240617386 & 3.6127 & 9&5607.9 $\pm$ 0.6&-&SP&53.24 $\pm$ 7.24&6.68 $\pm$ 0.91\\
  NS87 & 150.1509930 & 2.226192297 & 3.6782 & 3&5687.1 $\pm$ 0.7&5690.1 $\pm$ 1.3&RT&19.96 $\pm$ 3.92&2.62 $\pm$ 0.51\\
  NS88 & 150.1525301 & 2.238594726 & 4.5186 & 3&6708.8 $\pm$ 0.3&-&SP&7.91 $\pm$ 0.87&1.69 $\pm$ 0.19\\
  NS91 & 150.1864893 & 2.234161180 & 3.6828 & 9&5692.3 $\pm$ 0.9&-&SP&6.49 $\pm$ 1.89&0.86 $\pm$ 0.25\\
  NS92 & 150.1805399 & 2.235132627 & 3.8673 & 3&5917.2 $\pm$ 0.3&-&SP&12.74 $\pm$ 1.69&1.88 $\pm$ 0.25\\
  NS97 & 150.1916706 & 2.235040683 & 5.9697 & 9&8474.4 $\pm$ 1.2&-&SP&13.06 $\pm$ 2.93&5.37 $\pm$ 1.21\\
  NS98 & 150.1057122 & 2.328948835 & 3.2080 & 9&5116.1 $\pm$ 0.4&-&SP&11.11 $\pm$ 1.97&1.05 $\pm$ 0.19\\
  NS99 & 150.1192540 & 2.333246831 & 3.0062 & 9&4870.6 $\pm$ 0.4&-&SP&22.58 $\pm$ 3.04&1.82 $\pm$ 0.25\\
  NS100 & 150.1183659 & 2.321809003 & 3.2649 & 3&5184.9 $\pm$ 0.3&-&SP&19.95 $\pm$ 2.35&1.97 $\pm$ 0.23\\
  NS101 & 150.1247395 & 2.321276301 & 3.4198 & 9&5373.7 $\pm$ 0.6&-&SP&14.39 $\pm$ 2.53&1.58 $\pm$ 0.28\\
  NS102 & 150.1146837 & 2.323507763 & 3.4973 & 2&5469.1 $\pm$ 0.5&-&SP&13.88 $\pm$ 2.10&1.61 $\pm$ 0.24\\
  NS103 & 150.1152667 & 2.328748203 & 3.9439 & 9&6010.5 $\pm$ 0.3&-&SP&21.10 $\pm$ 2.37&3.27 $\pm$ 0.37\\
  NS105 & 150.1280974 & 2.320155687 & 4.4369 & 9&6609.6 $\pm$ 0.3&-&SP&11.86 $\pm$ 1.39&2.43 $\pm$ 0.29\\
  NS106 & 150.1137025 & 2.323197438 & 4.4971 & 9&6683.0 $\pm$ 0.4&-&SP&12.47 $\pm$ 1.84&2.64 $\pm$ 0.39\\
  NS107 & 150.1216946 & 2.320388333 & 4.5356 & 9&6729.5 $\pm$ 0.2&-&SP&9.03 $\pm$ 1.17&1.95 $\pm$ 0.25\\
\hline
\end{tabular}
\end{table*}
\begin{figure}
\centering
\includegraphics[scale=0.45]{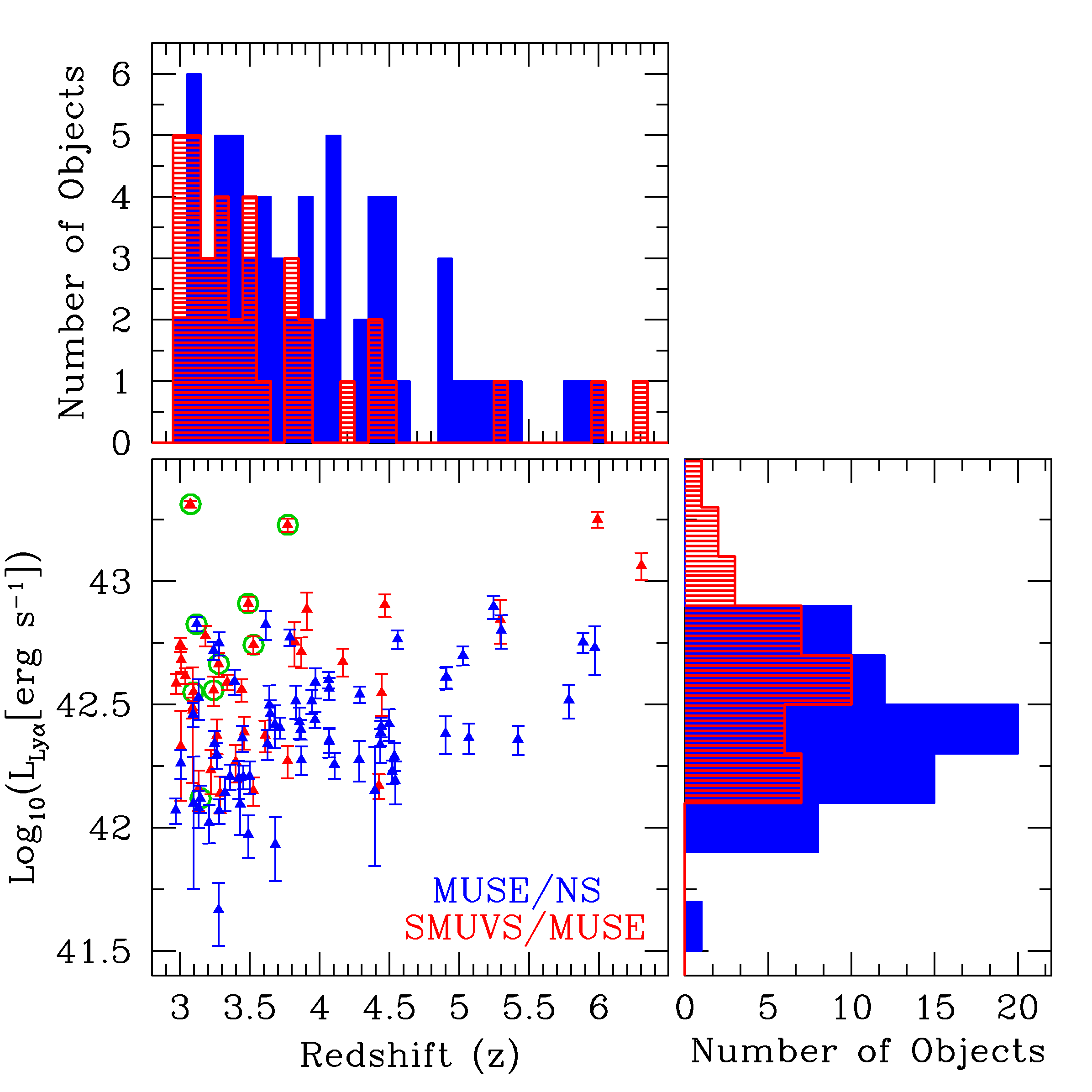}
\caption{{\it Upper panel:} Comparison in redshift distribution between the SMUVS sources (red) and the sources identified in the Blind search (blue). {\it Left panel:} Ly$\rm \alpha$ luminosity plotted as a function of redshift for our sources. The green circles identify the blue bump objects. {\it Right panel:} Histogram of the measured luminosities for both our samples. The blind search objects show slightly fainter luminosities compared to the SMUVS sources.}
\label{fig:L_LyA}
\end{figure}
Additionally to the Ly$\rm \alpha$ emitters in the SMUVS/MUSE sample, we also performed a blind search in the MUSE cubes and found 66 other sources presenting a secure or possible Ly$\rm \alpha$ emission ($\rm QF=2,3$ and 9). These new sources (named here MUSE/NS) have been identified by visually inspecting the MUSE data cubes and do not have a previous spectroscopic confirmation. We also verified that, in the specific framework of our analysis, performing an automated search in our cubes (i.e. by using the software LSDCat \citealt{2017A&A...602A.111H}), rather than a visual one, would not add new sources to our secure detections (QF $>$ 1). Since the MUSE/NS sample is not present in the SMUVS catalog, we have information on the redshift and measurements of the Ly$\rm \alpha$ flux and luminosity, but no estimate of the physical properties of these objects from SED fitting. The MUSE/NS sources and all their related quantities are listed in Table~\ref{table:noSMUVS}.\par
To make sure that these objects were truly a different population from the galaxies selected in SMUVS, we verified that none of these new sources was situated in masked areas of the survey. In fact, 61 of them were never selected for the catalog to begin with and 5 have a SMUVS neighbor within 1$''$, but are clearly different objects. We conclude that, since the SMUVS galaxies are detected based on a prior selection in the UltraVISTA $HKs$ stacks, these MUSE/NS sources are faint in the $HK_S$ stacks and/or in the images from the {\it Spitzer} $\rm 3.6\mu$m and $\rm 4.5\mu$m bands.\par
Given that the UltraVISTA images are deep, we can safely assume that the reason this objects are undetected is that they are below the mass limit of the survey. Even without having performed an SED fit, we can state that the MUSE/NS sample will likely have a very different mass distribution compared to the SMUVS/MUSE sample. This is likely the source of the discrepancies we find between the two populations. We put an upper limit on their stellar mass by citing the 50\% completeness limit reported in Table 1 of \citet{2018ApJ...864..166D}. For galaxies in $\rm 3.0 \le\, z \le\, 4.0$ the upper limit mass is $\rm log_{10}(M_*/M_\odot)=9.0$, for the range $\rm 4.0 <\, z \le\, 5.0$ it is 9.2, and finally for galaxies with $\rm 5.0 <\, z \le\, 6.0$ it is 9.4.\par
We measured the Ly$\rm \alpha$ emission and luminosity for the MUSE/NS sample in the same way as we did for the SMUVS/MUSE sources. The first difference we notice, is that the amount of blue bump profiles is strongly reduced compared to the SMUVS/MUSE sample (2/66 against 7/36, $\rm \sim 3$\% against $\rm \sim 19$\%). Furthermore, the single line profile appears in 44 spectra, constituting $\rm \sim 67$\% of the sample, which makes it more prominent than in the SMUVS/MUSE sample, where it was observed $\rm \sim 47$\% of the times. Finally, the red tail profile appears in 20 of our galaxies and is about as frequent as in the SMUVS/MUSE sample (MUSE/NS $\rm \sim 30$\%, SMUVS/MUSE $\rm \sim 31$\% ). If we compare SMUVS/MUSE to MUSE/NS both in line flux and luminosity using a KS test, we see that their distributions are different and that MUSE/NS are the fainter objects. This confirms the fact that blue bump profiles are harder to detect and thus the single line profiles increase in number for the fainter sources. Further evidence for this interpretation is that the blue bump objects are found in the brighter objects located at $\rm 3 \le z < 4$ of the SMUVS/MUSE (green circles in Fig.~\ref{fig:L_LyA}). The results of these KS tests can again be found in Table~\ref{table:KS}.\par
The right panel of Fig.~\ref{fig:L_LyA} illustrates this trend by showing the distribution of $\rm L_{Ly\alpha}$ for the two samples. We also plotted their distribution in redshift and how their luminosity evolves with it, in the upper and left panel in Fig.~\ref{fig:L_LyA} respectively. We see in the central panel of Fig.~\ref{fig:L_LyA} that for both samples the measured Ly$\rm \alpha$ luminosity plotted against redshift increases with it, as can be expected given there is a detection limit on the line flux. More interestingly, the MUSE/NS sources seem to lie close enough in redshift space to the SMUVS/MUSE sources, suggesting they could belong to the same physical structure. To test if this is the case, we defined a $\rm \Delta z$ dependent on redshift, such that two objects with said distance in redshift lie 2 Mpc apart. We then expanded our check also to the position on the sky and defined a sphere of 2 Mpc radius as our criterion to check for associations.\par
\begin{table}
\caption{Table showing the position and redshift of the three secure associations we identified using our method.}
\label{table:ASS}
\centering
\tiny
\begin{tabular}{|c|c|c c|c|}
\hline\hline
Association number & Obj-ID & RA & Dec & z \\
\hline
A1 & NS67 & 150.1634469 & 2.210140164 & 3.8656 \\
   & NS92 & 150.1805399 & 2.235132627 & 3.8673 \\
\hline
A2 & 76829 & 150.1926599 & 2.2198279 & 3.09 \\
   & 76877 & 150.1659591 & 2.2216971 & 3.0913 \\
\hline
A3 & NS4 & 150.0902794 & 2.208985301 & 4.0681 \\
   & NS75 & 150.094679 & 2.230024846 & 4.0667 \\
\hline
\end{tabular}
\end{table}
We find that we can identify 9 associations in the area covered by MUSE. If we restrict ourselves to only considering $\rm QF=3,4$ objects, then the number of associations we find drops to 3 (see Table.~\ref{table:ASS}). We compared the number of associations we find with MUSE in our small area to the number associations we could find in the part of the COSMOS field covered by SMUVS using spectroscopic confirmed objects known so far in the literature. We choose spectroscopic sources with $\rm QF=3,4$ only and apply the same criteria used on the MUSE data to identify associations. We find 16 associations over the entirety of the SMUVS/COSMOS field. If we assume that the MUSE rate in our small area is indicative of what we could find if MUSE covered SMUVS/COSMOS entirely, then we would expect to find $\rm \sim 360$ associations in this larger area. Unless cosmic variance is playing a big role in the area MUSE covered in our data, we estimate that MUSE has a $\rm \sim 20\times$ higher chance of detecting objects that could be physically linked, confirming the usefulness of MUSE for unbiased source detection.

%

\section{Summary and Conclusions}
\label{sec:conclu}
We made use of publicly available MUSE data in the COSMOS field to analyze a sample of 2997 photometrically selected galaxies from the SMUVS survey over an area of 20.79 $\rm arcmin^2$. We managed to detect and measure the redshift of 691 objects, of which 39 are located at $\rm z\ge2$. For these sources, we report two absorption line galaxies and one AGN in addition to 36 Lyman~$\rm \alpha$ emitters. Out of these 39 sources, all but one are new redshift measurements not previously present in the literature. Furthermore, we identify 66 additional Ly$\rm \alpha$ emitters by performing a blind search in the MUSE cubes. The values we measured for the Ly$\rm \alpha$ flux and luminosity of our sources are in line with other works using MUSE pointings with similar depth and conditions \citep{2017A&A...606A..12H,2019A&A...621A.107H}. We also detect three different line profiles in our combined Ly$\rm \alpha$ sample, hinting at different medium conditions in and around our objects. A quantitative analysis of these features was however out of the scope of this paper, whose goal was instead to investigate the differences in physical properties between the sources identified in the SMUVS catalog also detected in MUSE and those that could not be detected in MUSE.\par
The main result of our paper is to compare the physical properties of the SMUVS/MUSE and SMUVS/notMUSE sources. What we find is that, while there are some differences in the distribution of E(B-V), stellar mass and age in the two samples, their overall distribution does not vary substantially. We know that the SMUVS catalog is more sensitive to the brighter, more massive galaxies. MUSE instead has the only bias of being able to select objects that are at least bright enough in line emission to be detected. For $\rm z>2.9$ galaxies, this means they are bright Ly$\rm \alpha$ emitters, intense star-forming objects with little dust attenuation/HI absorption. Not finding a significant difference in physical properties between our two samples, even after applying a redshift cut and mass-matching them, could imply that the SMUVS selected Ly$\rm \alpha$ emitters we observe in MUSE are similar to the general population in SMUVS. Although SMUVS is not the largest nor deepest survey in COSMOS, it is the Spitzer survey with the largest area for its depth (only shallower than CANDELS, which covers an area 12 times smaller in COSMOS). We can thus probe a sample of galaxies that is more complete in parameter space than ever before.\par
After mass matching our samples, we see that SMUVS/MUSE is less obscured than SMUVS/notMUSE. We also notice how the age distribution of our Ly$\rm \alpha$ emitters shows a bimodality. Both these results have been found in other studies where Ly$\rm \alpha$ information has been combined with photometric SED fitting \citep{2008ApJ...674...70L,2009ApJ...691..465F,2009A&A...494..553P}. Furthermore, we have deepened our study of the bimodality in the age of our sample by studying the SFR of our objects qualitatively. We find that the younger objects are most likely experiencing their first burst of star formation. The older objects in our sample show instead a lower limit to their instantaneous SFR, derived from their Ly$\rm \alpha$ luminosity, that suggests they are undergoing a second burst of star formation, while the light of the galaxy is dominated by an older population of stars. This rejuvenation effect has also recently been observed in low-redshift objects by \citet{2019MNRAS.488L..99A} and \citet{2019arXiv190704337C}.\par
We also compare the SMUVS/MUSE and MUSE/NS samples. Using MUSE data allowed us an unbiased detection of all the sources that are bright in Ly$\rm \alpha$. In fact, we were able to perform a blind search in the MUSE cubes and enlarge our sample of Ly$\rm \alpha$ emitters by a factor of $\rm \sim 3$. MUSE confirms thus its potential for systematic searches of a given area in the sky. This sensitivity to emission lines also allowed us to detect 3 secure and previously undiscovered physical galaxy associations.\par
What is puzzling, however, is that MUSE detects MUSE/NS objects at luminosities around $10^{42}$ erg s$^{-1}$ and below, but does not detect such lower luminosities for the SMUVS/MUSE sample (see right histogram in Fig.~\ref{fig:L_LyA}). We determined that the MUSE/NS sample lies most likely below the mass completeness limit of the SMUVS survey. We note, however, that the Ly$\rm \alpha$ luminosities of the MUSE/NS sample have a significant overlap with the SMUVS/MUSE luminosities. Since the Ly$\rm \alpha$ line intensity is linked to the ionizing radiation emitted and not the mass of the object, this overlap is not surprising. It does, however, point out that we cannot attribute the lack of lower luminosity SMUVS/MUSE objects to a simple scaling effect due to the mass-selection in SMUVS. At the present, we do not have a definitive solution for this issue.\par
The area of improvement in this study is the depth of the pointings. Longer exposure times could yield a higher detection rate for the SMUVS selected sources (as of now, about 2/3 could not be detected) and deep enough spectra to measure a continuum. This would allow us to more accurately investigate the line shapes (we could distinguish P-Cygni profiles from single lines) and also measure equivalent widths for the brighter objects.
\begin{acknowledgements}
The authors would like to thank the anonymous referee for the helpful and constructing feedback. GR, GBC, KIC and SD acknowledge funding from the European Research Council through the award of the Consolidator Grant ID 681627-BUILDUP.\\
Based on observations collected at the European Southern Observatory under ESO Programme IDs 095.A-0240(A), 096.A-0090(A), 097.A-0160(A), 098.A-0017(A). Also based in part on observations carried out with the Spitzer Space Telescope, which is operated by the Jet Propulsion Laboratory, California Institute of Technology under a contract with NASA. This research made use of Astropy,\footnote{http://www.astropy.org} a community-developed core Python package for Astronomy \citep{astropy:2013,astropy:2018}.\\
http://adsabs.harvard.edu/abs/2018arXiv180102634T\\
http://adsabs.harvard.edu/abs/2013A\&A...558A..33A
\end{acknowledgements}

\bibliographystyle{aa} 
\bibliography{biblio}

\begin{thebibliography}{75}
\expandafter\ifx\csname natexlab\endcsname\relax\def\natexlab#1{#1}\fi

\bibitem[{{Acquaviva} {et~al.}(2012){Acquaviva}, {Vargas}, {Gawiser}, \&
  {Guaita}}]{2012ApJ...751L..26A}
{Acquaviva}, V., {Vargas}, C., {Gawiser}, E., \& {Guaita}, L. 2012, \apjl, 751,
  L26

\bibitem[{{Angthopo} {et~al.}(2019){Angthopo}, {Ferreras}, \&
  {Silk}}]{2019MNRAS.488L..99A}
{Angthopo}, J., {Ferreras}, I., \& {Silk}, J. 2019, \mnras, 488, L99

\bibitem[{{Arnouts} {et~al.}(1999){Arnouts}, {Cristiani}, {Moscardini},
  {Matarrese}, {Lucchin}, {Fontana}, \& {Giallongo}}]{1999MNRAS.310..540A}
{Arnouts}, S., {Cristiani}, S., {Moscardini}, L., {et~al.} 1999, \mnras, 310,
  540

\bibitem[{{Ashby} {et~al.}(2018){Ashby}, {Caputi}, {Cowley}, {Deshmukh},
  {Dunlop}, {Milvang-Jensen}, {Fynbo}, {Muzzin}, {McCracken}, {Le F{\`e}vre},
  {Huang}, \& {Zhang}}]{2018ApJS..237...39A}
{Ashby}, M.~L.~N., {Caputi}, K.~I., {Cowley}, W., {et~al.} 2018, \apjs, 237, 39

\bibitem[{{Astropy Collaboration} {et~al.}(2013){Astropy Collaboration},
  {Robitaille}, {Tollerud}, {Greenfield}, {Droettboom}, {Bray}, {Aldcroft},
  {Davis}, {Ginsburg}, {Price-Whelan}, {Kerzendorf}, {Conley}, {Crighton},
  {Barbary}, {Muna}, {Ferguson}, {Grollier}, {Parikh}, {Nair}, {Unther},
  {Deil}, {Woillez}, {Conseil}, {Kramer}, {Turner}, {Singer}, {Fox}, {Weaver},
  {Zabalza}, {Edwards}, {Azalee Bostroem}, {Burke}, {Casey}, {Crawford},
  {Dencheva}, {Ely}, {Jenness}, {Labrie}, {Lim}, {Pierfederici}, {Pontzen},
  {Ptak}, {Refsdal}, {Servillat}, \& {Streicher}}]{astropy:2013}
{Astropy Collaboration}, {Robitaille}, T.~P., {Tollerud}, E.~J., {et~al.} 2013,
  \aap, 558, A33

\bibitem[{{Bacon} {et~al.}(2010){Bacon}, {Accardo}, {Adjali}, {Anwand},
  {Bauer}, {Biswas}, {Blaizot}, {Boudon}, {Brau-Nogue}, {Brinchmann},
  {Caillier}, {Capoani}, {Carollo}, {Contini}, {Couderc}, {Daguis{\'e}},
  {Deiries}, {Delabre}, {Dreizler}, {Dubois}, {Dupieux}, {Dupuy}, {Emsellem},
  {Fechner}, {Fleischmann}, {Fran{\c c}ois}, {Gallou}, {Gharsa}, {Glindemann},
  {Gojak}, {Guiderdoni}, {Hansali}, {Hahn}, {Jarno}, {Kelz}, {Koehler},
  {Kosmalski}, {Laurent}, {Le Floch}, {Lilly}, {Lizon}, {Loupias}, {Manescau},
  {Monstein}, {Nicklas}, {Olaya}, {Pares}, {Pasquini}, {P{\'e}contal-Rousset},
  {Pell{\'o}}, {Petit}, {Popow}, {Reiss}, {Remillieux}, {Renault}, {Roth},
  {Rupprecht}, {Serre}, {Schaye}, {Soucail}, {Steinmetz}, {Streicher}, {Stuik},
  {Valentin}, {Vernet}, {Weilbacher}, {Wisotzki}, \&
  {Yerle}}]{2010SPIE.7735E..08B}
{Bacon}, R., {Accardo}, M., {Adjali}, L., {et~al.} 2010, in \procspie, Vol.
  7735, Ground-based and Airborne Instrumentation for Astronomy III, 773508

\bibitem[{{Bacon} {et~al.}(2016){Bacon}, {Piqueras}, {Conseil}, {Richard}, \&
  {Shepherd}}]{2016ascl.soft11003B}
{Bacon}, R., {Piqueras}, L., {Conseil}, S., {Richard}, J., \& {Shepherd}, M.
  2016, {MPDAF: MUSE Python Data Analysis Framework}, Astrophysics Source Code
  Library

\bibitem[{{Bertin} \& {Arnouts}(1996)}]{1996A&AS..117..393B}
{Bertin}, E. \& {Arnouts}, S. 1996, \aaps, 117, 393

\bibitem[{{Bridge} {et~al.}(2018){Bridge}, {Hayes}, {Melinder}, {{\"O}stlin},
  {Gronwall}, {Ciardullo}, {Atek}, {Cannon}, {Gronke}, {Guaita}, {Hagen},
  {Herenz}, {Kunth}, {Laursen}, {Mas-Hesse}, \& {Pardy}}]{2018ApJ...852....9B}
{Bridge}, J.~S., {Hayes}, M., {Melinder}, J., {et~al.} 2018, \apj, 852, 9

\bibitem[{{Caputi} {et~al.}(2011){Caputi}, {Cirasuolo}, {Dunlop}, {McLure},
  {Farrah}, \& {Almaini}}]{2011MNRAS.413..162C}
{Caputi}, K.~I., {Cirasuolo}, M., {Dunlop}, J.~S., {et~al.} 2011, \mnras, 413,
  162

\bibitem[{{Caputi} {et~al.}(2017){Caputi}, {Deshmukh}, {Ashby}, {Cowley},
  {Bisigello}, {Fazio}, {Fynbo}, {Le F{\`e}vre}, {Milvang-Jensen}, \&
  {Ilbert}}]{2017ApJ...849...45C}
{Caputi}, K.~I., {Deshmukh}, S., {Ashby}, M.~L.~N., {et~al.} 2017, \apj, 849,
  45

\bibitem[{{Cooke} {et~al.}(2019){Cooke}, {Kartaltepe}, {Tyler}, {Darvish},
  {Casey}, {Le F{\`e}vre}, {Salvato}, \& {Scoville}}]{2019arXiv190704337C}
{Cooke}, K.~C., {Kartaltepe}, J.~S., {Tyler}, K.~D., {et~al.} 2019, arXiv
  e-prints, arXiv:1907.04337

\bibitem[{{Cowley} {et~al.}(2018){Cowley}, {Caputi}, {Deshmukh}, {Ashby},
  {Fazio}, {Le F{\`e}vre}, {Fynbo}, {Ilbert}, {McCracken}, {Milvang-Jensen}, \&
  {Somerville}}]{2018ApJ...853...69C}
{Cowley}, W.~I., {Caputi}, K.~I., {Deshmukh}, S., {et~al.} 2018, \apj, 853, 69

\bibitem[{{Cowley} {et~al.}(2019){Cowley}, {Caputi}, {Deshmukh}, {Ashby},
  {Fazio}, {Le F{\`e}vre}, {Fynbo}, {Ilbert}, \&
  {Milvang-Jensen}}]{2019ApJ...874..114C}
{Cowley}, W.~I., {Caputi}, K.~I., {Deshmukh}, S., {et~al.} 2019, \apj, 874, 114

\bibitem[{{Curtis-Lake} {et~al.}(2013){Curtis-Lake}, {McLure}, {Dunlop},
  {Schenker}, {Rogers}, {Targett}, {Cirasuolo}, {Almaini}, {Ashby}, \&
  {Bradshaw}}]{2013MNRAS.429..302C}
{Curtis-Lake}, E., {McLure}, R.~J., {Dunlop}, J.~S., {et~al.} 2013, \mnras,
  429, 302

\bibitem[{{De Barros} {et~al.}(2017){De Barros}, {Pentericci}, {Vanzella},
  {Castellano}, {Fontana}, {Grazian}, {Conselice}, {Yan}, {Koekemoer}, \&
  {Cristiani}}]{2017A&A...608A.123D}
{De Barros}, S., {Pentericci}, L., {Vanzella}, E., {et~al.} 2017, \aap, 608,
  A123

\bibitem[{{Deshmukh} {et~al.}(2018){Deshmukh}, {Caputi}, {Ashby}, {Cowley},
  {McCracken}, {Fynbo}, {Le F{\`e}vre}, {Milvang-Jensen}, \&
  {Ilbert}}]{2018ApJ...864..166D}
{Deshmukh}, S., {Caputi}, K.~I., {Ashby}, M.~L.~N., {et~al.} 2018, \apj, 864,
  166

\bibitem[{{Diener} {et~al.}(2017){Diener}, {Wisotzki}, {Schmidt}, {Herenz},
  {Urrutia}, {Garel}, {Kerutt}, {Saust}, {Bacon}, {Cantalupo}, {Contini},
  {Guiderdoni}, {Marino}, {Richard}, {Schaye}, {Soucail}, \&
  {Weilbacher}}]{2017MNRAS.471.3186D}
{Diener}, C., {Wisotzki}, L., {Schmidt}, K.~B., {et~al.} 2017, \mnras, 471,
  3186

\bibitem[{{Dijkstra}(2017)}]{2017arXiv170403416D}
{Dijkstra}, M. 2017, ArXiv e-prints [\eprint[arXiv]{1704.03416}]

\bibitem[{{Erb} {et~al.}(2018){Erb}, {Steidel}, \&
  {Chen}}]{2018ApJ...862L..10E}
{Erb}, D.~K., {Steidel}, C.~C., \& {Chen}, Y. 2018, \apjl, 862, L10

\bibitem[{{Fazio} {et~al.}(2004){Fazio}, {Hora}, {Allen}, {Ashby}, {Barmby},
  {Deutsch}, {Huang}, {Kleiner}, {Marengo}, {Megeath}, {Melnick}, {Pahre},
  {Patten}, {Polizotti}, {Smith}, {Taylor}, {Wang}, {Willner}, {Hoffmann},
  {Pipher}, {Forrest}, {McMurty}, {McCreight}, {McKelvey}, {McMurray}, {Koch},
  {Moseley}, {Arendt}, {Mentzell}, {Marx}, {Losch}, {Mayman}, {Eichhorn},
  {Krebs}, {Jhabvala}, {Gezari}, {Fixsen}, {Flores}, {Shakoorzadeh}, {Jungo},
  {Hakun}, {Workman}, {Karpati}, {Kichak}, {Whitley}, {Mann}, {Tollestrup},
  {Eisenhardt}, {Stern}, {Gorjian}, {Bhattacharya}, {Carey}, {Nelson},
  {Glaccum}, {Lacy}, {Lowrance}, {Laine}, {Reach}, {Stauffer}, {Surace},
  {Wilson}, {Wright}, {Hoffman}, {Domingo}, \& {Cohen}}]{2004ApJS..154...10F}
{Fazio}, G.~G., {Hora}, J.~L., {Allen}, L.~E., {et~al.} 2004, \apjs, 154, 10

\bibitem[{{Finkelstein} {et~al.}(2009){Finkelstein}, {Rhoads}, {Malhotra}, \&
  {Grogin}}]{2009ApJ...691..465F}
{Finkelstein}, S.~L., {Rhoads}, J.~E., {Malhotra}, S., \& {Grogin}, N. 2009,
  \apj, 691, 465

\bibitem[{{Grogin} {et~al.}(2011){Grogin}, {Kocevski}, {Faber}, {Ferguson},
  {Koekemoer}, {Riess}, {Acquaviva}, {Alexander}, {Almaini}, {Ashby}, {Barden},
  {Bell}, {Bournaud}, {Brown}, {Caputi}, {Casertano}, {Cassata}, {Castellano},
  {Challis}, {Chary}, {Cheung}, {Cirasuolo}, {Conselice}, {Roshan Cooray},
  {Croton}, {Daddi}, {Dahlen}, {Dav{\'e}}, {de Mello}, {Dekel}, {Dickinson},
  {Dolch}, {Donley}, {Dunlop}, {Dutton}, {Elbaz}, {Fazio}, {Filippenko},
  {Finkelstein}, {Fontana}, {Gardner}, {Garnavich}, {Gawiser}, {Giavalisco},
  {Grazian}, {Guo}, {Hathi}, {H{\"a}ussler}, {Hopkins}, {Huang}, {Huang},
  {Jha}, {Kartaltepe}, {Kirshner}, {Koo}, {Lai}, {Lee}, {Li}, {Lotz}, {Lucas},
  {Madau}, {McCarthy}, {McGrath}, {McIntosh}, {McLure}, {Mobasher},
  {Moustakas}, {Mozena}, {Nandra}, {Newman}, {Niemi}, {Noeske}, {Papovich},
  {Pentericci}, {Pope}, {Primack}, {Rajan}, {Ravindranath}, {Reddy}, {Renzini},
  {Rix}, {Robaina}, {Rodney}, {Rosario}, {Rosati}, {Salimbeni}, {Scarlata},
  {Siana}, {Simard}, {Smidt}, {Somerville}, {Spinrad}, {Straughn}, {Strolger},
  {Telford}, {Teplitz}, {Trump}, {van der Wel}, {Villforth}, {Wechsler},
  {Weiner}, {Wiklind}, {Wild}, {Wilson}, {Wuyts}, {Yan}, \&
  {Yun}}]{2011ApJS..197...35G}
{Grogin}, N.~A., {Kocevski}, D.~D., {Faber}, S.~M., {et~al.} 2011, \apjs, 197,
  35

\bibitem[{{Gronke}(2017)}]{2017A&A...608A.139G}
{Gronke}, M. 2017, \aap, 608, A139

\bibitem[{{Guaita} {et~al.}(2011){Guaita}, {Acquaviva}, {Padilla}, {Gawiser},
  {Bond}, {Ciardullo}, {Treister}, {Kurczynski}, {Gronwall}, \&
  {Lira}}]{2011ApJ...733..114G}
{Guaita}, L., {Acquaviva}, V., {Padilla}, N., {et~al.} 2011, \apj, 733, 114

\bibitem[{{Gurung-Lopez} {et~al.}(2018){Gurung-Lopez}, {Orsi}, \&
  {Bonoli}}]{2018arXiv181109630G}
{Gurung-Lopez}, S., {Orsi}, A.~A., \& {Bonoli}, S. 2018, arXiv e-prints
  [\eprint[arXiv]{1811.09630}]

\bibitem[{{Hao} {et~al.}(2018){Hao}, {Huang}, {Xia}, {Zheng}, {Jiang}, \&
  {Li}}]{2018ApJ...864..145H}
{Hao}, C.-N., {Huang}, J.-S., {Xia}, X., {et~al.} 2018, \apj, 864, 145

\bibitem[{{Herenz} {et~al.}(2017){Herenz}, {Urrutia}, {Wisotzki}, {Kerutt},
  {Saust}, {Werhahn}, {Schmidt}, {Caruana}, {Diener}, {Bacon}, {Brinchmann},
  {Schaye}, {Maseda}, \& {Weilbacher}}]{2017A&A...606A..12H}
{Herenz}, E.~C., {Urrutia}, T., {Wisotzki}, L., {et~al.} 2017, \aap, 606, A12

\bibitem[{{Herenz} \& {Wisotzki}(2017)}]{2017A&A...602A.111H}
{Herenz}, E.~C. \& {Wisotzki}, L. 2017, \aap, 602, A111

\bibitem[{{Herenz} {et~al.}(2019){Herenz}, {Wisotzki}, {Saust}, {Kerutt},
  {Urrutia}, {Diener}, {Schmidt}, {Marino}, {de la Vieuville}, {Boogaard},
  {Schaye}, {Guiderdoni}, {Richard}, \& {Bacon}}]{2019A&A...621A.107H}
{Herenz}, E.~C., {Wisotzki}, L., {Saust}, R., {et~al.} 2019, \aap, 621, A107

\bibitem[{{Hern{\'a}n-Caballero} {et~al.}(2017){Hern{\'a}n-Caballero},
  {P{\'e}rez-Gonz{\'a}lez}, {Diego}, {Lagattuta}, {Richard}, {Schaerer},
  {Alonso-Herrero}, {Marino}, {Sklias}, {Alcalde Pampliega}, {Cava},
  {Conselice}, {Dannerbauer}, {Dom{\'{\i}}nguez-S{\'a}nchez}, {Eliche-Moral},
  {Esquej}, {Huertas-Company}, {Marques-Chaves}, {P{\'e}rez-Fournon}, {Rawle},
  {Rodr{\'{\i}}guez Espinosa}, {Rosa Gonz{\'a}lez}, \&
  {Rujopakarn}}]{2017ApJ...849...82H}
{Hern{\'a}n-Caballero}, A., {P{\'e}rez-Gonz{\'a}lez}, P.~G., {Diego}, J.~M.,
  {et~al.} 2017, \apj, 849, 82

\bibitem[{{Hu} {et~al.}(1998){Hu}, {Cowie}, \& {McMahon}}]{1998ApJ...502L..99H}
{Hu}, E.~M., {Cowie}, L.~L., \& {McMahon}, R.~G. 1998, \apjl, 502, L99

\bibitem[{{Ilbert} {et~al.}(2006){Ilbert}, {Arnouts}, {McCracken},
  {Bolzonella}, {Bertin}, {Le F{\`e}vre}, {Mellier}, {Zamorani}, {Pell{\`o}},
  {Iovino}, {Tresse}, {Le Brun}, {Bottini}, {Garilli}, {Maccagni}, {Picat},
  {Scaramella}, {Scodeggio}, {Vettolani}, {Zanichelli}, {Adami}, {Bardelli},
  {Cappi}, {Charlot}, {Ciliegi}, {Contini}, {Cucciati}, {Foucaud}, {Franzetti},
  {Gavignaud}, {Guzzo}, {Marano}, {Marinoni}, {Mazure}, {Meneux}, {Merighi},
  {Paltani}, {Pollo}, {Pozzetti}, {Radovich}, {Zucca}, {Bondi}, {Bongiorno},
  {Busarello}, {de La Torre}, {Gregorini}, {Lamareille}, {Mathez}, {Merluzzi},
  {Ripepi}, {Rizzo}, \& {Vergani}}]{2006A&A...457..841I}
{Ilbert}, O., {Arnouts}, S., {McCracken}, H.~J., {et~al.} 2006, \aap, 457, 841

\bibitem[{{Kakiichi} \& {Gronke}(2019)}]{2019arXiv190502480K}
{Kakiichi}, K. \& {Gronke}, M. 2019, arXiv e-prints, arXiv:1905.02480

\bibitem[{{Karman} {et~al.}(2017){Karman}, {Caputi}, {Caminha}, {Gronke},
  {Grillo}, {Balestra}, {Rosati}, {Vanzella}, {Coe}, {Dijkstra}, {Koekemoer},
  {McLeod}, {Mercurio}, \& {Nonino}}]{2017A&A...599A..28K}
{Karman}, W., {Caputi}, K.~I., {Caminha}, G.~B., {et~al.} 2017, \aap, 599, A28

\bibitem[{{Karman} {et~al.}(2014){Karman}, {Caputi}, {Trager}, {Almaini}, \&
  {Cirasuolo}}]{2014A&A...565A...5K}
{Karman}, W., {Caputi}, K.~I., {Trager}, S.~C., {Almaini}, O., \& {Cirasuolo},
  M. 2014, \aap, 565, A5

\bibitem[{{Kennicutt}(1998)}]{1998ARA&A..36..189K}
{Kennicutt}, Robert~C., J. 1998, \araa, 36, 189

\bibitem[{{Kimm} {et~al.}(2019){Kimm}, {Blaizot}, {Garel}, {Michel-Dansac},
  {Katz}, {Rosdahl}, {Verhamme}, \& {Haehnelt}}]{2019arXiv190105990K}
{Kimm}, T., {Blaizot}, J., {Garel}, T., {et~al.} 2019, arXiv e-prints
  [\eprint[arXiv]{1901.05990}]

\bibitem[{{Koekemoer} {et~al.}(2011){Koekemoer}, {Faber}, {Ferguson}, {Grogin},
  {Kocevski}, {Koo}, {Lai}, {Lotz}, {Lucas}, {McGrath}, {Ogaz}, {Rajan},
  {Riess}, {Rodney}, {Strolger}, {Casertano}, {Castellano}, {Dahlen},
  {Dickinson}, {Dolch}, {Fontana}, {Giavalisco}, {Grazian}, {Guo}, {Hathi},
  {Huang}, {van der Wel}, {Yan}, {Acquaviva}, {Alexander}, {Almaini}, {Ashby},
  {Barden}, {Bell}, {Bournaud}, {Brown}, {Caputi}, {Cassata}, {Challis},
  {Chary}, {Cheung}, {Cirasuolo}, {Conselice}, {Roshan Cooray}, {Croton},
  {Daddi}, {Dav{\'e}}, {de Mello}, {de Ravel}, {Dekel}, {Donley}, {Dunlop},
  {Dutton}, {Elbaz}, {Fazio}, {Filippenko}, {Finkelstein}, {Frazer}, {Gardner},
  {Garnavich}, {Gawiser}, {Gruetzbauch}, {Hartley}, {H{\"a}ussler},
  {Herrington}, {Hopkins}, {Huang}, {Jha}, {Johnson}, {Kartaltepe},
  {Khostovan}, {Kirshner}, {Lani}, {Lee}, {Li}, {Madau}, {McCarthy},
  {McIntosh}, {McLure}, {McPartland}, {Mobasher}, {Moreira}, {Mortlock},
  {Moustakas}, {Mozena}, {Nandra}, {Newman}, {Nielsen}, {Niemi}, {Noeske},
  {Papovich}, {Pentericci}, {Pope}, {Primack}, {Ravindranath}, {Reddy},
  {Renzini}, {Rix}, {Robaina}, {Rosario}, {Rosati}, {Salimbeni}, {Scarlata},
  {Siana}, {Simard}, {Smidt}, {Snyder}, {Somerville}, {Spinrad}, {Straughn},
  {Telford}, {Teplitz}, {Trump}, {Vargas}, {Villforth}, {Wagner}, {Wandro},
  {Wechsler}, {Weiner}, {Wiklind}, {Wild}, {Wilson}, {Wuyts}, \&
  {Yun}}]{2011ApJS..197...36K}
{Koekemoer}, A.~M., {Faber}, S.~M., {Ferguson}, H.~C., {et~al.} 2011, \apjs,
  197, 36

\bibitem[{{Kova{\v c}} {et~al.}(2010){Kova{\v c}}, {Lilly}, {Cucciati},
  {Porciani}, {Iovino}, {Zamorani}, {Oesch}, {Bolzonella}, {Knobel},
  {Finoguenov}, {Peng}, {Carollo}, {Pozzetti}, {Caputi}, {Silverman}, {Tasca},
  {Scodeggio}, {Vergani}, {Scoville}, {Capak}, {Contini}, {Kneib}, {Le
  F{\`e}vre}, {Mainieri}, {Renzini}, {Bardelli}, {Bongiorno}, {Coppa}, {de la
  Torre}, {de Ravel}, {Franzetti}, {Garilli}, {Guzzo}, {Kampczyk},
  {Lamareille}, {Le Borgne}, {Le Brun}, {Maier}, {Mignoli}, {Pello}, {Perez
  Montero}, {Ricciardelli}, {Tanaka}, {Tresse}, {Zucca}, {Abbas}, {Bottini},
  {Cappi}, {Cassata}, {Cimatti}, {Fumana}, {Koekemoer}, {Maccagni}, {Marinoni},
  {McCracken}, {Memeo}, {Meneux}, \& {Scaramella}}]{2010ApJ...708..505K}
{Kova{\v c}}, K., {Lilly}, S.~J., {Cucciati}, O., {et~al.} 2010, \apj, 708, 505

\bibitem[{{Kriek} {et~al.}(2015){Kriek}, {Shapley}, {Reddy}, {Siana}, {Coil},
  {Mobasher}, {Freeman}, {de Groot}, {Price}, {Sanders}, {Shivaei}, {Brammer},
  {Momcheva}, {Skelton}, {van Dokkum}, {Whitaker}, {Aird}, {Azadi}, {Kassis},
  {Bullock}, {Conroy}, {Dav{\'e}}, {Kere{\v s}}, \&
  {Krumholz}}]{2015ApJS..218...15K}
{Kriek}, M., {Shapley}, A.~E., {Reddy}, N.~A., {et~al.} 2015, \apjs, 218, 15

\bibitem[{{Lai} {et~al.}(2008){Lai}, {Huang}, {Fazio}, {Gawiser}, {Ciardullo},
  {Damen}, {Franx}, {Gronwall}, {Labbe}, \& {Magdis}}]{2008ApJ...674...70L}
{Lai}, K., {Huang}, J.-S., {Fazio}, G., {et~al.} 2008, \apj, 674, 70

\bibitem[{{Le F{\`e}vre} {et~al.}(2005){Le F{\`e}vre}, {Vettolani}, {Garilli},
  {Tresse}, {Bottini}, {Le Brun}, {Maccagni}, {Picat}, {Scaramella},
  {Scodeggio}, {Zanichelli}, {Adami}, {Arnaboldi}, {Arnouts}, {Bardelli},
  {Bolzonella}, {Cappi}, {Charlot}, {Ciliegi}, {Contini}, {Foucaud},
  {Franzetti}, {Gavignaud}, {Guzzo}, {Ilbert}, {Iovino}, {McCracken}, {Marano},
  {Marinoni}, {Mathez}, {Mazure}, {Meneux}, {Merighi}, {Paltani}, {Pell{\`o}},
  {Pollo}, {Pozzetti}, {Radovich}, {Zamorani}, {Zucca}, {Bondi}, {Bongiorno},
  {Busarello}, {Lamareille}, {Mellier}, {Merluzzi}, {Ripepi}, \&
  {Rizzo}}]{2005A&A...439..845L}
{Le F{\`e}vre}, O., {Vettolani}, G., {Garilli}, B., {et~al.} 2005, \aap, 439,
  845

\bibitem[{{Lilly} {et~al.}(2007){Lilly}, {Le F{\`e}vre}, {Renzini}, {Zamorani},
  {Scodeggio}, {Contini}, {Carollo}, {Hasinger}, {Kneib}, {Iovino}, {Le Brun},
  {Maier}, {Mainieri}, {Mignoli}, {Silverman}, {Tasca}, {Bolzonella},
  {Bongiorno}, {Bottini}, {Capak}, {Caputi}, {Cimatti}, {Cucciati}, {Daddi},
  {Feldmann}, {Franzetti}, {Garilli}, {Guzzo}, {Ilbert}, {Kampczyk}, {Kovac},
  {Lamareille}, {Leauthaud}, {Le Borgne}, {McCracken}, {Marinoni}, {Pello},
  {Ricciardelli}, {Scarlata}, {Vergani}, {Sanders}, {Schinnerer}, {Scoville},
  {Taniguchi}, {Arnouts}, {Aussel}, {Bardelli}, {Brusa}, {Cappi}, {Ciliegi},
  {Finoguenov}, {Foucaud}, {Franceschini}, {Halliday}, {Impey}, {Knobel},
  {Koekemoer}, {Kurk}, {Maccagni}, {Maddox}, {Marano}, {Marconi}, {Meneux},
  {Mobasher}, {Moreau}, {Peacock}, {Porciani}, {Pozzetti}, {Scaramella},
  {Schiminovich}, {Shopbell}, {Smail}, {Thompson}, {Tresse}, {Vettolani},
  {Zanichelli}, \& {Zucca}}]{2007ApJS..172...70L}
{Lilly}, S.~J., {Le F{\`e}vre}, O., {Renzini}, A., {et~al.} 2007, \apjs, 172,
  70

\bibitem[{{Mallery} {et~al.}(2012){Mallery}, {Mobasher}, {Capak}, {Kakazu},
  {Masters}, {Ilbert}, {Hemmati}, {Scarlata}, {Salvato}, \&
  {McCracken}}]{2012ApJ...760..128M}
{Mallery}, R.~P., {Mobasher}, B., {Capak}, P., {et~al.} 2012, \apj, 760, 128

\bibitem[{{Marchi} {et~al.}(2019){Marchi}, {Pentericci}, {Guaita}, {Talia},
  {Castellano}, {Hathi}, {Schaerer}, {Amorin}, {Carnall}, {Charlot},
  {Chevallard}, {Cullen}, {Finkelstein}, {Fontana}, {Fontanot}, {Garilli},
  {Hibon}, {Koekemoer}, {Maccagni}, {McLure}, {Papovich}, {Pozzetti}, \&
  {Saxena}}]{2019arXiv190308593M}
{Marchi}, F., {Pentericci}, L., {Guaita}, L., {et~al.} 2019, arXiv e-prints
  [\eprint[arXiv]{1903.08593}]

\bibitem[{{Martin} {et~al.}(2015){Martin}, {Dijkstra}, {Henry}, {Soto},
  {Danforth}, \& {Wong}}]{2015ApJ...803....6M}
{Martin}, C.~L., {Dijkstra}, M., {Henry}, A., {et~al.} 2015, \apj, 803, 6

\bibitem[{{Martinache} {et~al.}(2018){Martinache}, {Rettura}, {Dole},
  {Lehnert}, {Frye}, {Altieri}, {Beelen}, {B{\'e}thermin}, {Le Floc'h},
  {Giard}, {Hurier}, {Lagache}, {Montier}, {Omont}, {Pointecouteau},
  {Polletta}, {Puget}, {Scott}, {Soucail}, \& {Welikala}}]{2018A&A...620A.198M}
{Martinache}, C., {Rettura}, A., {Dole}, H., {et~al.} 2018, \aap, 620, A198

\bibitem[{{Mas-Hesse} {et~al.}(2003){Mas-Hesse}, {Kunth}, {Tenorio-Tagle},
  {Leitherer}, {Terlevich}, \& {Terlevich}}]{2003ApJ...598..858M}
{Mas-Hesse}, J.~M., {Kunth}, D., {Tenorio-Tagle}, G., {et~al.} 2003, \apj, 598,
  858

\bibitem[{{McCracken} {et~al.}(2012){McCracken}, {Milvang-Jensen}, {Dunlop},
  {Franx}, {Fynbo}, {Le F{\`e}vre}, {Holt}, {Caputi}, {Goranova}, {Buitrago},
  {Emerson}, {Freudling}, {Hudelot}, {L{\'o}pez-Sanjuan}, {Magnard}, {Mellier},
  {M{\o}ller}, {Nilsson}, {Sutherland}, {Tasca}, \&
  {Zabl}}]{2012A&A...544A.156M}
{McCracken}, H.~J., {Milvang-Jensen}, B., {Dunlop}, J., {et~al.} 2012, \aap,
  544, A156

\bibitem[{{McLinden} {et~al.}(2014){McLinden}, {Rhoads}, {Malhotra},
  {Finkelstein}, {Richardson}, {Smith}, \& {Tilvi}}]{2014MNRAS.439..446M}
{McLinden}, E.~M., {Rhoads}, J.~E., {Malhotra}, S., {et~al.} 2014, \mnras, 439,
  446

\bibitem[{{Nakajima} {et~al.}(2018){Nakajima}, {Fletcher}, {Ellis},
  {Robertson}, \& {Iwata}}]{2018MNRAS.477.2098N}
{Nakajima}, K., {Fletcher}, T., {Ellis}, R.~S., {Robertson}, B.~E., \& {Iwata},
  I. 2018, \mnras, 477, 2098

\bibitem[{{Ono} {et~al.}(2010){Ono}, {Ouchi}, {Shimasaku}, {Akiyama}, {Dunlop},
  {Farrah}, {Lee}, {McLure}, {Okamura}, \& {Yoshida}}]{2010MNRAS.402.1580O}
{Ono}, Y., {Ouchi}, M., {Shimasaku}, K., {et~al.} 2010, \mnras, 402, 1580

\bibitem[{{Orlitov{\'a}} {et~al.}(2018){Orlitov{\'a}}, {Verhamme}, {Henry},
  {Scarlata}, {Jaskot}, {Oey}, \& {Schaerer}}]{2018A&A...616A..60O}
{Orlitov{\'a}}, I., {Verhamme}, A., {Henry}, A., {et~al.} 2018, \aap, 616, A60

\bibitem[{{Pentericci} {et~al.}(2009){Pentericci}, {Grazian}, {Fontana},
  {Castellano}, {Giallongo}, {Salimbeni}, \& {Santini}}]{2009A&A...494..553P}
{Pentericci}, L., {Grazian}, A., {Fontana}, A., {et~al.} 2009, \aap, 494, 553

\bibitem[{{Pentericci} {et~al.}(2010){Pentericci}, {Grazian}, {Scarlata},
  {Fontana}, {Castellano}, {Giallongo}, \& {Vanzella}}]{2010A&A...514A..64P}
{Pentericci}, L., {Grazian}, A., {Scarlata}, C., {et~al.} 2010, \aap, 514, A64

\bibitem[{{Piqueras} {et~al.}(2017){Piqueras}, {Conseil}, {Shepherd}, {Bacon},
  {Leclercq}, \& {Richard}}]{2017arXiv171003554P}
{Piqueras}, L., {Conseil}, S., {Shepherd}, M., {et~al.} 2017, ArXiv e-prints
  [\eprint[arXiv]{1710.03554}]

\bibitem[{{Price-Whelan} {et~al.}(2018){Price-Whelan}, {Sip{\H{o}}cz},
  {G{\"u}nther}, {Lim}, {Crawford}, {Conseil}, {Shupe}, {Craig}, {Dencheva},
  {Ginsburg}, {VanderPlas}, {Bradley}, {P{\'e}rez-Su{\'a}rez}, {de Val-Borro},
  {Paper Contributors}, {Aldcroft}, {Cruz}, {Robitaille}, {Tollerud},
  {Coordination Committee}, {Ardelean}, {Babej}, {Bach}, {Bachetti}, {Bakanov},
  {Bamford}, {Barentsen}, {Barmby}, {Baumbach}, {Berry}, {Biscani}, {Boquien},
  {Bostroem}, {Bouma}, {Brammer}, {Bray}, {Breytenbach}, {Buddelmeijer},
  {Burke}, {Calderone}, {Cano Rodr{\'\i}guez}, {Cara}, {Cardoso}, {Cheedella},
  {Copin}, {Corrales}, {Crichton}, {D{\textquoteright}Avella}, {Deil},
  {Depagne}, {Dietrich}, {Donath}, {Droettboom}, {Earl}, {Erben}, {Fabbro},
  {Ferreira}, {Finethy}, {Fox}, {Garrison}, {Gibbons}, {Goldstein}, {Gommers},
  {Greco}, {Greenfield}, {Groener}, {Grollier}, {Hagen}, {Hirst}, {Homeier},
  {Horton}, {Hosseinzadeh}, {Hu}, {Hunkeler}, {Ivezi{\'c}}, {Jain}, {Jenness},
  {Kanarek}, {Kendrew}, {Kern}, {Kerzendorf}, {Khvalko}, {King}, {Kirkby},
  {Kulkarni}, {Kumar}, {Lee}, {Lenz}, {Littlefair}, {Ma}, {Macleod},
  {Mastropietro}, {McCully}, {Montagnac}, {Morris}, {Mueller}, {Mumford},
  {Muna}, {Murphy}, {Nelson}, {Nguyen}, {Ninan}, {N{\"o}the}, {Ogaz}, {Oh},
  {Parejko}, {Parley}, {Pascual}, {Patil}, {Patil}, {Plunkett}, {Prochaska},
  {Rastogi}, {Reddy Janga}, {Sabater}, {Sakurikar}, {Seifert}, {Sherbert},
  {Sherwood-Taylor}, {Shih}, {Sick}, {Silbiger}, {Singanamalla}, {Singer},
  {Sladen}, {Sooley}, {Sornarajah}, {Streicher}, {Teuben}, {Thomas},
  {Tremblay}, {Turner}, {Terr{\'o}n}, {van Kerkwijk}, {de la Vega}, {Watkins},
  {Weaver}, {Whitmore}, {Woillez}, {Zabalza}, \& {Contributors}}]{astropy:2018}
{Price-Whelan}, A.~M., {Sip{\H{o}}cz}, B.~M., {G{\"u}nther}, H.~M., {et~al.}
  2018, \aj, 156, 123

\bibitem[{{Remolina-Guti{\'e}rrez} \&
  {Forero-Romero}(2019)}]{2019MNRAS.482.4553R}
{Remolina-Guti{\'e}rrez}, M.~C. \& {Forero-Romero}, J.~E. 2019, \mnras, 482,
  4553

\bibitem[{{Scoville} {et~al.}(2007){Scoville}, {Aussel}, {Brusa}, {Capak},
  {Carollo}, {Elvis}, {Giavalisco}, {Guzzo}, {Hasinger}, {Impey}, {Kneib},
  {LeFevre}, {Lilly}, {Mobasher}, {Renzini}, {Rich}, {Sanders}, {Schinnerer},
  {Schminovich}, {Shopbell}, {Taniguchi}, \& {Tyson}}]{2007ApJS..172....1S}
{Scoville}, N., {Aussel}, H., {Brusa}, M., {et~al.} 2007, \apjs, 172, 1

\bibitem[{{Shapley} {et~al.}(2003){Shapley}, {Steidel}, {Pettini}, \&
  {Adelberger}}]{2003ApJ...588...65S}
{Shapley}, A.~E., {Steidel}, C.~C., {Pettini}, M., \& {Adelberger}, K.~L. 2003,
  \apj, 588, 65

\bibitem[{{Smith} {et~al.}(2019){Smith}, {Ma}, {Bromm}, {Finkelstein},
  {Hopkins}, {Faucher-Gigu{\`e}re}, \& {Kere{\v s}}}]{2019MNRAS.484...39S}
{Smith}, A., {Ma}, X., {Bromm}, V., {et~al.} 2019, \mnras, 484, 39

\bibitem[{{Sobral} {et~al.}(2018){Sobral}, {Matthee}, {Darvish}, {Smail},
  {Best}, {Alegre}, {R{\"o}ttgering}, {Mobasher}, {Paulino-Afonso}, {Stroe}, \&
  {Oteo}}]{2018MNRAS.477.2817S}
{Sobral}, D., {Matthee}, J., {Darvish}, B., {et~al.} 2018, \mnras, 477, 2817

\bibitem[{{Soto} {et~al.}(2016){Soto}, {Lilly}, {Bacon}, {Richard}, \&
  {Conseil}}]{2016MNRAS.458.3210S}
{Soto}, K.~T., {Lilly}, S.~J., {Bacon}, R., {Richard}, J., \& {Conseil}, S.
  2016, \mnras, 458, 3210

\bibitem[{{Taniguchi} {et~al.}(2007){Taniguchi}, {Scoville}, {Murayama},
  {Sanders}, {Mobasher}, {Aussel}, {Capak}, {Ajiki}, {Miyazaki}, {Komiyama},
  {Shioya}, {Nagao}, {Sasaki}, {Koda}, {Carilli}, {Giavalisco}, {Guzzo},
  {Hasinger}, {Impey}, {LeFevre}, {Lilly}, {Renzini}, {Rich}, {Schinnerer},
  {Shopbell}, {Kaifu}, {Karoji}, {Arimoto}, {Okamura}, \&
  {Ohta}}]{2007ApJS..172....9T}
{Taniguchi}, Y., {Scoville}, N., {Murayama}, T., {et~al.} 2007, \apjs, 172, 9

\bibitem[{{Tenorio-Tagle} {et~al.}(1999){Tenorio-Tagle}, {Silich}, {Kunth},
  {Terlevich}, \& {Terlevich}}]{1999MNRAS.309..332T}
{Tenorio-Tagle}, G., {Silich}, S.~A., {Kunth}, D., {Terlevich}, E., \&
  {Terlevich}, R. 1999, \mnras, 309, 332

\bibitem[{{Vanzella} {et~al.}(2018){Vanzella}, {Nonino}, {Cupani},
  {Castellano}, {Sani}, {Mignoli}, {Calura}, {Meneghetti}, {Gilli}, {Comastri},
  {Mercurio}, {Caminha}, {Caputi}, {Rosati}, {Grillo}, {Cristiani}, {Balestra},
  {Fontana}, \& {Giavalisco}}]{2018MNRAS.476L..15V}
{Vanzella}, E., {Nonino}, M., {Cupani}, G., {et~al.} 2018, \mnras, 476, L15

\bibitem[{{Verhamme} {et~al.}(2018){Verhamme}, {Garel}, {Ventou}, {Contini},
  {Bouch{\'e}}, {Herenz}, {Richard}, {Bacon}, {Schmidt}, {Maseda}, {Marino},
  {Brinchmann}, {Cantalupo}, {Caruana}, {Cl{\'e}ment}, {Diener}, {Drake},
  {Hashimoto}, {Inami}, {Kerutt}, {Kollatschny}, {Leclercq}, {Patr{\'{\i}}cio},
  {Schaye}, {Wisotzki}, \& {Zabl}}]{2018MNRAS.478L..60V}
{Verhamme}, A., {Garel}, T., {Ventou}, E., {et~al.} 2018, \mnras, 478, L60

\bibitem[{{Verhamme} {et~al.}(2008){Verhamme}, {Schaerer}, {Atek}, \&
  {Tapken}}]{2008A&A...491...89V}
{Verhamme}, A., {Schaerer}, D., {Atek}, H., \& {Tapken}, C. 2008, \aap, 491, 89

\bibitem[{{Weilbacher} {et~al.}(2006){Weilbacher}, {Roth},
  {P{\'e}contal-Rousset}, \& {Bacon}}]{2006NewAR..50..405W}
{Weilbacher}, P.~M., {Roth}, M.~M., {P{\'e}contal-Rousset}, A., \& {Bacon}, R.
  2006, \nar, 50, 405

\bibitem[{{Weilbacher} {et~al.}(2012){Weilbacher}, {Streicher}, {Urrutia},
  {Jarno}, {P{\'e}contal-Rousset}, {Bacon}, \&
  {B{\"o}hm}}]{2012SPIE.8451E..0BW}
{Weilbacher}, P.~M., {Streicher}, O., {Urrutia}, T., {et~al.} 2012, in
  \procspie, Vol. 8451, Software and Cyberinfrastructure for Astronomy II,
  84510B

\bibitem[{{Weilbacher} {et~al.}(2014){Weilbacher}, {Streicher}, {Urrutia},
  {P{\'e}contal-Rousset}, {Jarno}, \& {Bacon}}]{2014ASPC..485..451W}
{Weilbacher}, P.~M., {Streicher}, O., {Urrutia}, T., {et~al.} 2014, in
  Astronomical Society of the Pacific Conference Series, Vol. 485, Astronomical
  Data Analysis Software and Systems XXIII, ed. N.~{Manset} \& P.~{Forshay},
  451

\bibitem[{{Werner} {et~al.}(2004){Werner}, {Roellig}, {Low}, {Rieke}, {Rieke},
  {Hoffmann}, {Young}, {Houck}, {Brandl}, {Fazio}, {Hora}, {Gehrz}, {Helou},
  {Soifer}, {Stauffer}, {Keene}, {Eisenhardt}, {Gallagher}, {Gautier}, {Irace},
  {Lawrence}, {Simmons}, {Van Cleve}, {Jura}, {Wright}, \&
  {Cruikshank}}]{2004ApJS..154....1W}
{Werner}, M.~W., {Roellig}, T.~L., {Low}, F.~J., {et~al.} 2004, \apjs, 154, 1

\bibitem[{{Yuma} {et~al.}(2010){Yuma}, {Ohta}, {Yabe}, {Shimasaku}, {Yoshida},
  {Ouchi}, {Iwata}, \& {Sawicki}}]{2010ApJ...720.1016Y}
{Yuma}, S., {Ohta}, K., {Yabe}, K., {et~al.} 2010, \apj, 720, 1016

\bibitem[{{Zhang} {et~al.}(2019){Zhang}, {Ouchi}, {Itoh}, {Shibuya}, {Ono},
  {Harikane}, {Inoue}, {Rauch}, {Kikuchihara}, {Nakajima}, {Yajima}, {Arata},
  {Abe}, {Iwata}, {Kashikawa}, {Kawanomoto}, {Kikuta}, {Kobayashi}, {Kusakabe},
  {Mawatari}, {Nagao}, {Shimasaku}, \& {Taniguchi}}]{2019arXiv190509841Z}
{Zhang}, H., {Ouchi}, M., {Itoh}, R., {et~al.} 2019, arXiv e-prints,
  arXiv:1905.09841

\end{thebibliography}
%

\end{document}